\documentclass[12pt,letter]{article}
\usepackage{graphicx, epsfig, color}
\newcommand{\be}{\begin{equation}}
\newcommand{\ee}{\end{equation}}
\newcommand{\bea}{\begin{eqnarray}}
\newcommand{\eea}{\end{eqnarray}}

\newcommand{\bphik}{$B_d \to\phi K_S$ }

\newcommand{\sQ}{{\tilde{Q}}}
\newcommand{\sq}{{\tilde{q}}}
\newcommand{\sU}{{\tilde{U}}}
\newcommand{\sD}{{\tilde{D}}}

\begin{document}
\begin{titlepage}
\begin{flushright}
MCTP-04-41\\
MADPH-04-1392 \\
hep-ph/0407351
\end{flushright}
\vspace{0.5cm}
\begin{center}
{\Large \bf High Scale Study of Possible \bphik CP Physics} \\
\vspace{1cm} \renewcommand{\thefootnote}{\fnsymbol{footnote}}
{\large Gordon L. Kane$^1$\footnote[1]{Email: gkane@umich.edu}, 
 Haibin Wang$^2$\footnote[2]{Email: haibinw@physics.purdue.edu}, 
 Lian-Tao Wang$^3$\footnote[3]{Email: liantaow@pheno.physics.wisc.edu}, 
 Ting T. Wang$^1$\footnote[4]{Email: tingwang@umich.edu}} \\
\vspace{1.cm} \renewcommand{\thefootnote}{\arabic{footnote}}
{\it 
1. Michigan Center for Theoretical Physics, Ann
  Arbor, MI 48109, USA \\
2. Department of Physics, Purdue University, West Lafayette, IN 47906, USA \\
3. Department of Physics, University of Wisconsin, Madison, WI 53705, USA 
}
\end{center}
\vspace{0.5cm}

\begin{abstract}
Some rare decay processes are particularly sensitive to physics beyond the
Standard Model (SM) because they have no SM tree contributions. We focus
on one of these, $B_{d}\rightarrow \phi K_{s}$. Our study is in terms of
the high scale effective theory, and high scale models for the
underlying theory, while previous studies have been focusing on the low
scale effective Lagrangian. We examine phenomenologically the high scale
parameter space with full calculations, but largely report the results
in terms of mass insertion techniques since they are then easily
pictured. We also determine the ranges of
different mass insertions that could produce large non-SM CP effects.  Then
we exhibit classes of high scale models that can or cannot provide large
non-SM CP effects, thus demonstrating that data on $B_{d}\rightarrow \phi
K_{s}$ can probe both supersymmetry breaking and the underlying high scale
theory and even make relatively direct contact with string-motivated models.
We provide a novel and systematic technique to understand the relations
between high and low scale parameters from RGE running. We include all
constraints from other data, particularly $b\rightarrow s\gamma$ and EDMs.
\end{abstract}
\vspace*{1cm}
\end{titlepage}

\section{Introduction}

CP violation is a fascinating subject in particle physics. Complex
parameters in the Lagrangian could give rise to numerous interesting
observables in low energy experiments. Many, especially those associated
with the third generation of quarks, are either not well measured or
controversial. They represent new opportunities to discover new physics and
uncover new flavor physics in the near future. 
If possible FCNC parameters are small, 
some presently unknown mechanism or symmetry has to be found to
explain why they are small. Sizable new CP violations, as well as
the KM phase itself, probe fundamental flavor physics, which is
closely related to supersymmetry breaking and string theory.

In the Standard Model, the KM phase $\delta _{KM}$ and the strong phase $%
\theta _{QCD}$ are the only sources for CP violation. The bounds on the
electric dipole moment(EDM) of the neutron imply that $\theta _{QCD}<10^{-9}$%
, so we will ignore it for b physics. Therefore, effectively the only source
of CP violation in the SM is the non-zero $\delta _{KM}$. Many
experiments have been analyzed to measure this phase and consistent results
would increasingly establish the KM mechanism of CP violation. Before the
measurement of the time dependent CP asymmetry in the process of $%
B_{d}\rightarrow \phi K_{S}$ , all experiments indeed gave a consistent
measure of $\delta_{KM}=60^{\circ}\pm 14^\circ$ \cite{Eidelman}.
On the other hand, many proposals for new physics beyond the Standard Model
allow new sources of CP violation. Some of them are constrained by the
experimental bounds from electric dipole moments and measurements in Kaon
physics. Others could give rise to potentially interesting deviations.

Low energy supersymmetry is the most compelling candidate for the new physics 
\cite{Chung:2003fi}. The most general soft supersymmetry breaking Lagrangian
carries a significant number of phases which could be interesting new
sources of the CP violation. Therefore, it is important to study the ways in
which CP violation in the soft supersymmetry breaking Lagrangian could show
up in the experiments.

Since processes involving the superpartners (hence the new parameters)
necessarily occur at loop level, the new CP violating effects could only be
significant in the processes where SM contributions are also suppressed.
Important examples of such processes include $b\rightarrow s \gamma$ and $%
B_d \rightarrow \phi K_S$.

It is interesting to notice that recently there is a potential discrepancy
between the CP asymmetry measured in the process $B_{d}\rightarrow \phi
K_{S} $ (denoted by $S_{\phi K_{S}}$) and the SM prediction:
\begin{eqnarray}
S_{\phi K} \simeq S_{\psi K} \simeq \sin 2\beta = 0.736
\end{eqnarray}
The latest results from BaBar \cite{Aubert:2002nx, Browder} is 
\begin{equation}
S_{\phi K_{S}}=0.47\pm 0.34_{-0.06}^{+0.08},  \label{flBaBar}
\end{equation}
and from BELLE \cite{Abe:2002bx, Browder} is 
\begin{equation}
S_{\phi K_{S}}=-0.96\pm 0.50_{-0.11}^{+0.09}.  \label{flBELLE}
\end{equation}
Because this process is one of a few that are unusually sensitive to CP
physics beyond the SM, it is worthwhile studying it in detail however the
experimental situation is finally resolved. And of course if a deviation
from the SM prediction is confirmed it is exceptionally important.

Since $\delta _{KM}$ is the only source of CP violation in the SM, if
a deviation from the SM prediction for $S_{\phi K}$
is confirmed, it will be a clear sign of new physics beyond
the SM and the new physics must be relevant to the weak scale. The
superpartners in the loops must be at a mass scale that guarantees their
production at the Tevatron and LHC. In the Minimal Supersymmetric Standard
Model(MSSM), there are 42 new CP violation phases in addition to $\delta
_{KM}$ and $\theta _{QCD}$. Although many of these phases are already
constrained by various experiments \cite{Gabbiani:1996hi}, there are still
plenty of them which are not strongly constrained. It was shown in 
\cite{Khalil:2002fm,Kane:2003zi,Harnik:2002vs,Baek:2003kb,Chakraverty:2003uv}, 
that by tuning some of these not strongly
constrained CP violation phases at the weak scale, BELLE's result could
be described without violating other experimental bounds.

The most stringent contraints on CP violation arise from several electric
dipole moment measurements. Since we are mostly interested in CP violation
associated with the last generations of quarks, the most relevant constraint
will come from the mercury EDM which depends on the chromo-magnetic operator
of the strange quark (see detailed discussion in Sec.~\ref{constraints}).
From the point of view of a general low energy MSSM, this should not give a
direct constraint on CP violation observed in flavor changing B decays since
the EDM is only sensitive to flavor diagonal phases. However, we have to
bear in mind that the soft Lagrangian is actually defined at some high scale
at which supersymmetry is broken. Generically, we expect soft
parameters at that scale to carry large phases that can give large
deviations from the Standard Model in CP violating observables. The
renormalization group running mixes the flavor-diagonal and flavor
off-diagonal parameters and the mixing is enhanced by the large Logs in the
conventional picture of gauge coupling unification. One of the main results
of this paper is the study of this effect and its implications for the high
energy allowed parameter space.

With better understanding of the allowed high energy parameter space, we then
study and constrain some models of flavor structures. Previous studies
have focused on weak scale phenomenology. Here instead we emphasize the
properties of the high scale theory and their relation to the data. We
do both a phenomenological analysis of what high scale properties would
be needed to explain a deviation from the SM, and an examination of what
kinds of high scale underlying theories could or could not have the
needed properties.

This paper is organized as follows. 
We describes the most relevant CP violation observables and constraints
in Section 2. A model independent study of the input high scale parameter space
is performed in Section 3, utilizing both semi-analytic and numerical
methods to study the RGE effects. With the knowledge obtained from the model
independent study, we examine several classes of models of
high scale flavor structure in Section 4. We present our conclusions in
Section 5 and a detailed semi-analytic study of MSSM RGE effects using a
new approach in the appendix, which
provides a clear picture of the interplays among the flavor parameters.

\section{CP Violation Observables and Constraints}

\label{constraints} In this section, we review the main experimental
contraints and prospects of detecting CP violation beyond the
Standard Model. We discuss the Mercury EDM and $b\rightarrow s\gamma $ as
they usually provide the most useful constraints. We use $B_{d}\rightarrow
\phi K_{S}$ as the main example of an observable potential deviation.
Other constraints \cite{Gabbiani:1996hi} involving first or second 
families of quarks are less important for our purpose. We will not
discuss them here, but still require our results to satisfy those experimental
bounds. We emphasize that, although mass insertions generally provide a clearer 
picture of the flavor changing, we make use of the full squark mass matrix 
in our calculation. The necessity is evident in the discussions of
the importance of various multiple-mass-insertions in the following.

\subsection{Mercury Electric Dipole Moment}

The EDM bound on Hg puts constraints on the CP violation phases of the soft
terms. It requires the strange quark chromo EDM (CEDM) 
to satisfy \cite{Romalis:2000mg, Falk:1999tm}
\begin{equation}
|e d_{s}^{C}|<5.8\times 10^{-25}e\mbox{cm},  \label{fl:EDMbd}
\end{equation}
assuming vanishing up and down quarks CEDM.
However, there are significant theoretical hardronic uncertainties in
extracting $d_{s}^{C}$ from ${\rm Hg^{199}}$
\footnote{In our study, we therefore allow the theoretical
calculation to have a factor of 3 uncertainty.}. 

In the MSSM, although chargino exchange diagrams can
give sizeable contributions to the strange quark CEDM \cite{Endo:2003te},
the contribution is usually dominated by the gluino
exchange diagrams 
\begin{equation}
e d_{s}^{C}=ce\frac{\alpha _{s}}{4\pi }\frac{m_{\tilde{g}}}{m_{\tilde{q}}^{2}}
\mbox{Im}(\Delta _{22}^{d,LR})\times L(x)+L\leftrightarrow R,
\label{edmDelta}
\end{equation} 
with $c=0.91$ \cite{Demir:2003js}. $L(x)$, where $x=m_{\tilde{q}%
}^{2}/m_{\tilde{g}}^{2}$, is a loop function (or its appropriate
derivative). Notice it is different from $c=3.3$ usually quoted in the
literature \cite{Arnowitt:1990eh}. This difference comes from the different
definition of the chromo-dipole operator. The chromo-dipole operator used
in obtaining eq.(~\ref{edmDelta}) is defined as 
$ \frac{i}{2} g_s \bar{q}_i t^a G^a_{\mu \nu} \sigma^{\mu \nu} \gamma_5
q_i$
which includes the strong coupling $g_{s}$.
With this definition, which is also used in a recent study 
\cite{Hisano:2003iw}, there should be no large scaling of this operator.

$\Delta _{ij}^{f,LR}, ~(f=u, d)$ is a generic mixing parameter between 
left-handed $i$-th generation and
right-handed $j$-th generation of up- or down-type squarks.
It could involve single or
multiple mass insertion parameters which are directly defined from the soft
parameters. Several examples are 
\begin{eqnarray}
\Delta _{22}^{d,LR} &=&(\delta _{LR}^{d})_{22},  \nonumber \\
\Delta _{22}^{d,LR} &=&\frac{1}{2!}(\delta _{LL}^{d})_{23}(\delta
_{LR}^{d})_{32},\ \ \frac{1}{2!}(\delta _{LR}^{d})_{23}(\delta
_{RR}^{d})_{23}^{\ast },  \nonumber \\
\Delta _{22}^{d,LR} &=&\frac{1}{3!}(\delta _{LR}^{d})_{23}(\delta
_{LR}^{d})_{33}^{\ast }(\delta _{LR}^{d})_{32},\ \ \frac{1}{3!}(\delta
_{LL}^{d})_{23}(\delta _{LR}^{d})_{33}(\delta _{RR}^{d})_{32}
\end{eqnarray}%
In the cases of double mass insertion, there is an extra factor of $1/2$
coming from the Taylor expansion. Although flavor changing parameters do not
contribute to the EDM at leading order, they do enter at the next order
through combinations with other flavor parameters. A particular triple mass
insertion is studied in Ref~\cite{Hisano:2003iw}. Using Eq.~\ref{edmDelta},
we obtain for that combination 
\begin{equation}
ed_{s}^{C}=ce\frac{\alpha _{s}}{4\pi }\frac{m_{\tilde{g}}}{m_{\tilde{q}}^{2}}%
\left( -\frac{11}{30}\right) \frac{1}{3!}\mbox{Im}((\delta^d_{LL})_{23}
(\delta^d_{LR})_{33} (\delta^d_{RR})_{32}),  \label{edmtridelta}
\end{equation}
where we have set gluino and average squark masses to be equal,
$\tilde{m}_g = \tilde{m}_q$, when evaluating the loop function.
Notice also the extra $1/6$ coming from the Taylor expansion. Combining
this factor with the the scaling behavior, we found that the constraint of
Ref.~\cite{Hisano:2003iw} should be about a factor of 20 less restrictive
than they report.  \footnote{%
In a subsequent study \cite{Hisano:2004tf}, the same group of authors
changed their result after private communication from us. Their later
results agreed with the results presented in this paper.}.
Nevertheless, this combination will still provide some
constraint on the high energy CP violating parameters.

\subsection{$b \rightarrow s \protect\gamma$}
$b\to s\gamma$ is a process where the
Standard Model tree level contribution is absent.
Therefore, the SUSY contribution, which enters at
one-loop order, could be comparable to the Standard Model processes. As a
result, new flavor parameters in the soft Lagrangian will be significantly
constrained by this process. We use the following bound in our calculations: 
\begin{equation}
2.0\times 10^{-4}<\mbox{BR}(b\rightarrow s\gamma )<4.5\times 10^{-4}
\end{equation}

In the MSSM, the SUSY contribution to BR($b\rightarrow s\gamma $) is usually
assumed to be dominated by the chargino loop contribution to the $O_{7\gamma
}$(or $O_{7\gamma }^{\prime }$) operator:
\begin{eqnarray}
C_{7\gamma }^{\mbox{New}} &\propto &g_{2}^{2}\frac{m_{b}\tan \beta }{m_{W}}%
V_{12}\Delta _{23}^{u,LL}\times L_{\tilde{h}-\tilde{W}}  \nonumber \\
&+&g_{2}^{2}\lambda ^{2}\frac{m_{t}m_{b}\tan \beta }{m_{W}^{2}}\Delta
_{33}^{u,LR}V_{22}\times L_{\tilde{h}}  \nonumber \\
&+&g_{2}^{2}\frac{m_{b}}{M_{2}}X_{23}V_{11}\times L_{\tilde{W}}.
\label{bsgammac7}
\end{eqnarray}%
where $X_{23}$ is either $\Delta _{23}^{u,LL}$ or $\lambda ^{2}$. $V_{ij}$
are chargino mixing matrix elements. The three different lines correspond to
Higgsino-Wino, Higgsino, and Wino loops, respectively. Typically, the
Higgsino-Wino loop gives the dominant contribution. The others could be
important in some circumstances as well. Eq.~\ref{bsgammac7} is schematic
and designed to show the dependence on various flavor parameters. We have
not shown explicitly the charged Higgs contribution as it does not depend on
flavor changing soft parameters. There is a similar expression for the
chirality flipped $O_{7}^{\prime }$ operator as well \cite{Everett:2001yy}.

$\Delta _{ij}^{u,AB}$s should again be understood as compounded parameters.
In this paper, we are mainly interested in the flavor physics parameters
associated with the down sector. However, as we will argue in section 4, 
some of them are related to the up-sector
flavor parameter via $SU(2)$ gauge symmetry and CKM mixings. Therefore, 
$Br(b\rightarrow s\gamma)$ will provide important constraints in those cases.
On the other hand, the current experiments do not impose strong
constraints on the CP asymmetry of $b\rightarrow s\gamma$. So we will not
discuss it here.


\subsection{$B_d \rightarrow \phi K_S$}


The time-dependent CP asymmetry measured in the process of $B_d \to\phi K_S$
or $B_d\to J/\psi K_S$ is expressed in the following way: 
\begin{eqnarray}
a_{f}(t)&=&\frac{\Gamma(\overline B^0_d(t)\to f)-\Gamma(B^0_d (t)\to f)} {%
\Gamma(\overline B^0_d(t)\to f)+\Gamma(B^0_d (t)\to f)} \\
&=&C_f \cos\Delta M_{B_d} t +S_f \sin\Delta M_{B_d} t  \nonumber
\label{fltdcp}
\end{eqnarray}
where $f=\phi K_S\mbox{ or } J/\Phi K_S$ depending on which process we are
studying. 
Defining as usual 
\begin{equation}
\lambda_f \equiv \left( \frac{q}{p} \right) \frac{\bar{A}_f}{A_f},
\end{equation}
we have 
\begin{equation}
C_f = \frac{1-|\lambda_f|^2}{1+|\lambda_f|^2}, \ \ S_f = \frac{2Im(\lambda_f).%
}{1+|\lambda_f|^2}
\end{equation}
In the Standard Model, we have 
\begin{eqnarray}
\lambda_{\phi K_S}&=&\left(\frac{q}{p}\right) \frac{\overline A_{\phi K_S}%
}{A_{\phi K_S}} \\
&=&-\frac{V_{td}}{V_{td}^*}\frac{V_{ts}^*}{V_{ts}} \frac{V_{cs} V_{cd}^*}{%
V_{cs}^* V_{cd}}  \nonumber
\end{eqnarray}
This is invariant under redefinitions of phases.

The SUSY contributions to $\lambda _{f}$ can also be written in a phase
rotation invariant fashion. To simplify the formula, we assume SUSY 
contributions only
modify the decay amplitude significantly, not the mixing part.
We obtain
\begin{equation}
\lambda _{\phi K_{S}}=-e^{-i2\beta }\frac{1+r_{23}e^{i\theta
_{23}}+r_{32}e^{i\theta _{32}}}{1+r_{23}e^{-i\theta _{23}}+r_{32}e^{-i\theta
_{32}}},  \label{fllambdaSUSY}
\end{equation}
where we have defined 
\begin{equation}
r_{23}e^{i\theta _{23}}=\frac{b_{23}}{a}\frac{(\Delta _{23}^{d,LR})}{%
V_{ts}^{\ast }V_{tb}}\frac{M_{3}}{|M_{3}|},\ \ r_{32}e^{i\theta _{32}}=\frac{%
b_{32}}{a}\frac{(\Delta _{32}^{d,LR})^{\ast }}{V_{ts}^{\ast }V_{tb}}\frac{%
M_{3}^{\ast }}{|M_{3}|},  \label{fltheta32}
\end{equation}
in order to distinguish the $O_{8g}$ and $O_{8g}^{\prime }$ contributions.
$\frac{b_{23}}{a}$ is the ratio of the magnitudes of the SUSY and SM
contributions.
$\theta _{23}$ and $\theta _{32}$ are explicitly phase-rotation invariant.
They are two of the 42 new CP violation phases in the
MSSM\cite{Lebedev:2002wq}. In other words,
a non-zero $\theta _{23}$ or $\theta _{32}$ characterizes CP violation
beyond the KM mechanism. From eq.(\ref{fllambdaSUSY}),
it's clear that when $\theta _{23}$ and $\theta _{32}$ are both zero, the
time dependent CP asymmetry measured in the B factories should be 
\begin{equation}
S_{\phi K}=\sin 2\beta =0.736
\end{equation}%
and thus confirm the standard model KM mechanism of CP violation. Any
deviation from this result would imply a new CP violation source other than $%
\delta _{KM}$. In the framework of the MSSM, it would imply that some CP
violation phase such as $\theta _{23}$ or $\theta _{32}$ is non-zero. Notice
that although the SM CP violation phase $\beta $ and the new phase(s) $%
\theta _{23}$(or $\theta _{32}$) are independent parameters from the low
energy effective theory point of view, it's still possible that these phases
are correlated in fundamental flavor physics. Generically, the
number of independent phases in the fundamental theory could be less than
the number of phases carried by low energy parameters. 

In the Wolfenstein parameterization, $V_{ts}$ and $V_{tb}$ are real. In the
MSSM, the gluino phase can be rotated to zero by using the $U(1)_{R}$ symmetry.
Then $\theta _{23}$ and $\theta _{32}$ are just the phases of $\Delta
_{23}^{d,LR}$ and $\Delta _{32}^{d,LR}$, respectively. In the following, we
will use this parameterization, but one should keep in mind that when we say
there is a non-vanishing phase of these two MIs, what we really mean is the
phase of the combined quantity in eq.(\ref{fltheta32}).

It was shown in \cite{Kane:2003zi} that if at the weak scale, 
\footnote{%
Hereafter, when necessary we will use $(...)|_{W}$ and $(...)|_{\Lambda }$
to denote the weak scale and GUT scale values, respectively.} 
\begin{equation}
\Delta _{23}^{d,LR}|_{W}\mbox{ or }\Delta _{32}^{d,LR}|_{W}\sim {\mathcal{O}}%
(10^{-2})\times e^{i\phi },  \label{wnlr}
\end{equation}%
where $\phi $ is a non-trivial phase, then $S_{\phi K}$ can deviate
significantly from its SM value.
We will follow Ref. \cite{Kane:2003zi} to calculate the hadronic
matrix elements, using the BBNS method \cite{Beneke:1999br}.
In this approach, the strong phases arise from four classes of diagrams,
vertex corrections, penguins, hard scattering with spectator quarks and
annihilation diagrams.
It is important to properly account for the power corrections originated
from the latter two classes of diagrams.
These contributions should be subleading in the BBNS factorization, but
they involve infrared divergent integrals. To regularize them,
we follow BBNS and parameterize the integrals as 
$\Delta = (1+\rho e^{i\phi}) \log(m_B/\lambda_h)$ with 
$\lambda_h = 500 ~{\rm MeV}$.
This parameterization introduces hadronic uncertainties into our calculation.
However, as discussed in greater details in \cite{Kane:2003zi},
the uncertainties 
affect the branching ratios more than the CP asymmetries, and we focus
on the CP asymmetries in the present paper. The uncertainties also
decrease quickly for heavier gluino masses if we assume the
validity of BBNS factorization and choose moderate values of $\rho$.
We henceforth set $\rho = 0$ since our gluino will be heavy. In the present
paper we also mainly compare CP asymmetries for different models, and we
expect relative results to be even less sensitive to hadronic physics.

In this paper, we would like to study high scale models which can give large
non-SM CP violation in $B_{d}\rightarrow \phi K_{S}$ . We will use these as
examples of classes of models which could give rise to interesting low
energy CP violation beyond the SM. One interesting result is that some
classes of high scale theories do not allow one to obtain results such as
eq. (\ref{wnlr}) without violating other constraints.
Thus the low scale data can rather directly probe high
scale theories.

\section{Model Independent Study of High Energy Parameter Space}

\label{modelindepsection}

\subsection{RGE Running of High Energy FCNC Parameters}

Supersymmetry is used to stabilize the large hierarchy between the GUT scale
and weak scale. Therefore, the soft parameters at the input scale, which is
assumed to be close to the GUT scale, are related to those in the low energy
effective soft Lagrangian only after the inclusion of radiative corrections
enhanced by large logarithms. For CP violation in FCNC processes it is
crucial to take those RGE running effects into account. In this section, we
briefly summarize the results which are very useful for qualitatively
understanding the constraints on the high energy parameter space (see the
Appendix for a detailed study).

As explained in the Appendix, a systematic and novel way to study the RGE
running of the flavor parameters is to use what we call the High Energy
SuperCKM (HES) basis. In the HES basis, we are dealing with approximate
physical parameters and enjoying the presence of several small parameters,
such as $\lambda \sim 0.22$. The qualitative result of that study is
summarized in Table~\ref{rgedeltatable}.
\begin{table}[h]
\begin{center}
\begin{tabular}{|c|c|c|}
\hline
Parameter & Universal Contribution & Feeds into \\ \hline
$(\delta^d_{LL})_{23}$ & $\sim \eta \lambda^2 y_t^2 \sim 0.01$ & $-$ \\ \hline
$(\delta^d_{RR})_{23}$ & $\sim \eta^2 y_b^2 y_t^2 \lambda^2 < 10^{-4}$ & $-$
\\ \hline
$(\delta^d_{LR})_{23}$ & $\sim \frac{m_b}{m_{{\tilde{q}}}} \eta \lambda^2
y_t^2 < 10^{-4}$ & \ \ $(\delta^d_{LL})_{23} \sim 50 (\delta^d_{LR})^*_{23}
(\delta^d_{LR})_{33} $ \ \  \\ \hline
$(\delta^d_{LR})_{32}$ & \ \ $\sim \frac{m_b}{m_{{\tilde{q}}}} \eta^2 y_b
y_t^2 \lambda^2 < 10^{-5}$ \ \  & \ \ $(\delta^d_{RR})_{23}\sim 100
(\delta^d_{LR})_{32} (\delta^d_{LR})^*_{33}$ \ \  \\ \hline
\end{tabular}%
\end{center}
\caption{RGE Analysis of High Energy FCNC Parameters. $\protect\eta \sim
|t_{EW}-t_{GUT}|/16\protect\pi ^{2}\sim 0.2$ is the loop integration
parameter. }
\label{rgedeltatable}
\end{table}

Notice that starting from a
universal boundary condition at the high scale, the FCNC parameters still
acquire non-zero values due to small CKM mixing effects enhanced by the RGE
running. We call them universal contributions to the flavor parameters.
Important effects of this type are summarized in the second column of Table~%
\ref{rgedeltatable}. They should be regarded as important \textquotedblleft
model-independent\textquotedblright\ values of those parameters\footnote{%
Of cause, one could start with a set of non-zero FCNC parameters in such a
way that their initial values exactly cancel the RGE generated contribution.
This would require a conspiracy between the high energy fundamental flavor
physics and the radiative corrections associated with lower energy scales.
We do not consider such a possibility.}.

It is also important to notice that different FCNC parameters generically 
\textit{mix} through the RGE running. In particular, starting with one FCNC
parameter, others will be generated through the mixing. We document some of
the most important mixing effects in the third column of Table~\ref%
{rgedeltatable}. One of the obvious features is the \textquotedblleft
decoupling\textquotedblright\ bbehavior of LL and RR flavor parameters. Being
dimension two parameters, they would not enter the running of the
trilinears. As discussed in the appendix, the mixings between LL and RR
flavor parameters are also suppressed either by second generation quark
masses or by small CKM mixings.

Using the results of the RGE study, we now examine the connection of the
high energy FCNC soft parameters and the low energy observables. We
again remark that while we present results for simplicity i terms of
mass insertions, we actually do full numerical analyses without using
the mass insertion approximation.

\subsection{High Scale Parameter Space}

We focus on
the three observables discussed in the previous section. In a generic flavor
model, all of the flavor parameters could be non-zero at high scale.
However, in order to illustrate the constraints effectively, we study the
cases in which only one of those parameters is non-zero at a time.
The scenario is then referred to by the sole non-vanishing 
input scale flavor parameter, {\it e.g.},
$(\delta^d_{LL})_{23}|_{\Lambda_{GUT}}$
scenario which implies that $(\delta^d_{LL})_{23}$ is the only non-vanishing
mass insertion at input scale $\Lambda_{GUT}$.
We will identify the input scale as the GUT scale, where the gaugino
masses are
\begin{equation}
M_1 = M_2 = M_3 = 300 ~GeV,
\end{equation}
unless otherwise noted. The overall scale of the
squark mass matrices $m_0$ and trilinear terms are set to be $300 ~GeV$ 
as well. We will choose $\tan \beta = 15$ at weak scale throughout this section.

\subsubsection{$(\protect\delta^d_{LL})_{23}|_{\Lambda_{GUT}}$}

\begin{figure}[h!]
\begin{center}
\epsfig{file=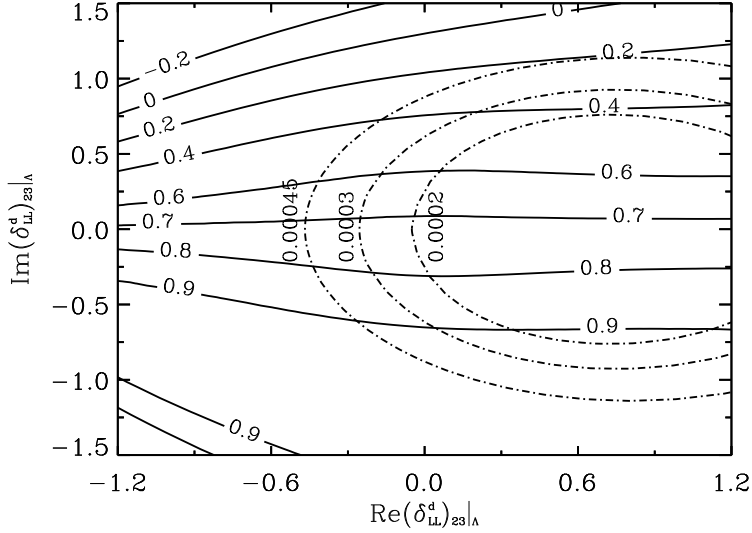, height=12cm,angle=90}
\caption{The correlation of $S_{\phi K_S}$ (solid lines) and 
BR$(b\to s\gamma)$
(dashed lines) is shown in this plot. The $x$- and $y$- axises are the
real and imaginary parts of $(\delta^d_{LL})_{23}$ at the input scale,
respectively.
The CEDM of the $s$ quark receives relatively small contributions in this
scenario and imposes no constraint.
Here we set $M_2 = M_3$ at the input scale.
\label{LLregion}
}
\end{center}
\end{figure}

\begin{figure}[h!]
\begin{center}
\epsfig{file=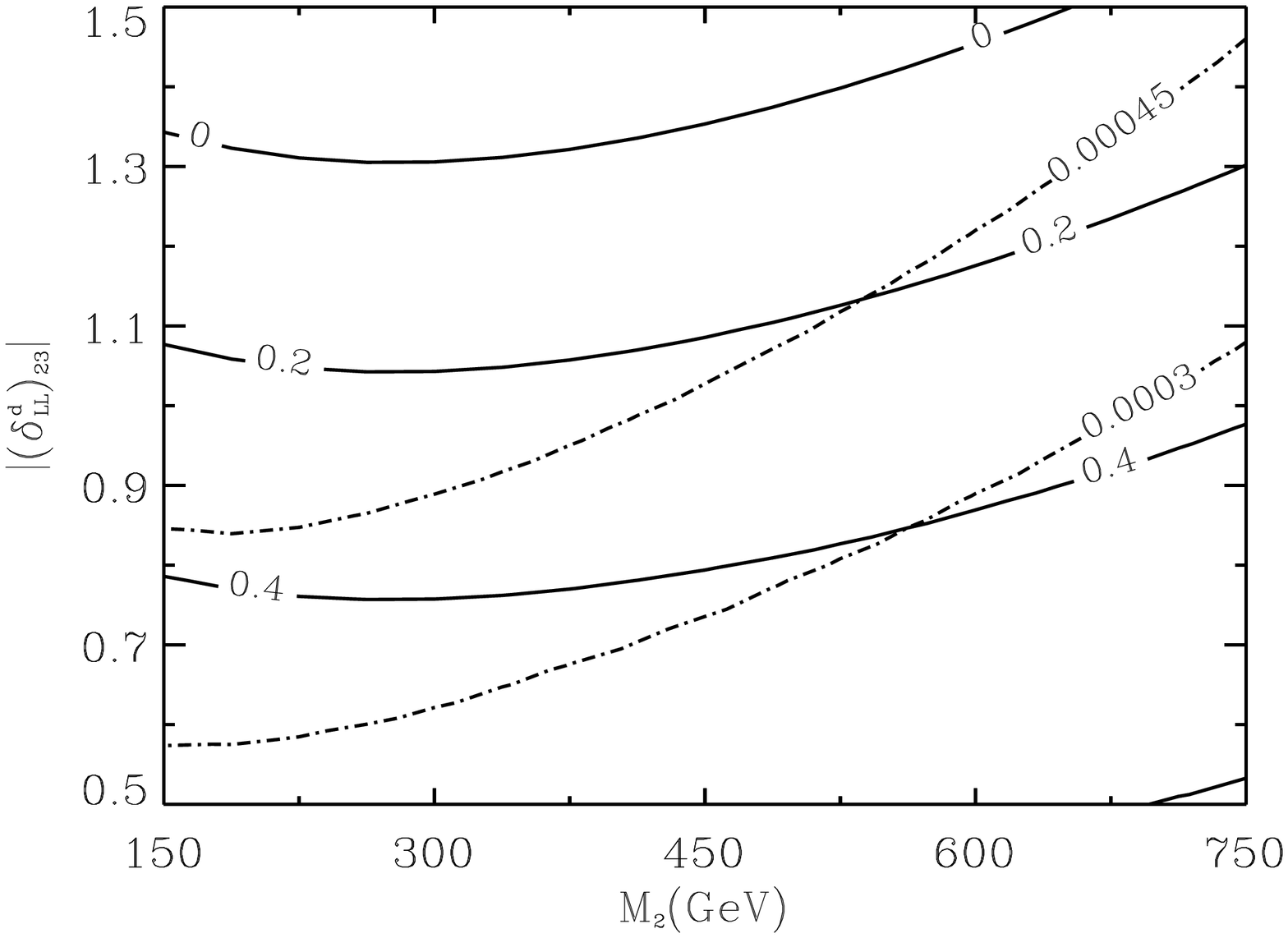,height=12cm, angle=90}
\caption{The correlation of $S_{\phi K_S}$ (solid lines) and 
BR$(b\to s\gamma)$
(dashed lines) is shown in this plot.
Here in contrast to Fig. \ref{LLregion},
we scan $M_2$(in GeV) and $|(\delta^d_{LL})_{23}|$
with $M_3=300$GeV and $Arg((\delta^d_{LL})_{23})=\pi/2$ fixed at the
input scale. \label{LLregion-M2}
}
\end{center}
\end{figure}

$(\delta _{LL}^{d})_{23}$ could contribute to $B_{d}\rightarrow \phi K_{S}$
through the chromo-dipole operator $O_{8g}$
\begin{equation}
O_{8g} = - \frac{g_s}{8 \pi^2}m_b \bar{s} \sigma_{\mu \nu} (1+\gamma_5) 
G^{\mu \nu} b
\end{equation}
with $C_{8g}\propto (\delta _{LL}^{d})_{23}(\delta _{LR}^{d})_{33}$. 
If it had a
large CP violating phase it could give rise to a large deviation from the SM
value of the CP asymmetry of $B_{d}\rightarrow \phi K_{S}$ .

Unlike many other cases, this CP violating parameter is not 
constrained by the EDM bound at the high scale. The reason is as
follows. The lowest order
contribution of $(\delta _{LL}^{d})_{23}$ to the mercury EDM is proportional
to $(\delta _{LR}^{d})_{32}^{\ast }(\delta _{LL}^{d})_{23}^{\ast }$. Using
the result of Table~\ref{rgedeltatable}, we see that starting with only $%
(\delta _{LL}^{d})_{23}$ at the high scale would not generate $(\delta
_{LR}^{d})_{32}$ through RGE running. Therefore, this type of contribution
to the EDM is suppressed. The next order contribution is through
the combination $(\delta _{RR}^{d})_{23}(\delta _{LR}^{d})_{33}^{\ast
}(\delta _{LL}^{d})_{23}^{\ast }\propto (\delta _{RR}^{d})_{23}C_{8g}^{\ast
}$. As indicated in Table~\ref{rgedeltatable}, the universal contributions
to $(\delta _{RR}^{d})_{23}$ due to RGE runnings is small. Adding the fact
that LL and RR FCNC soft parameters do not mix significantly under the RGE
running, this type of contribution is also suppressed. To summarize, the
scenario in which $(\delta _{LL}^{d})_{23}$ is the only flavor off-diagonal
complex soft parameter at the high scale is almost unconstrained by low
energy EDM measurements. Notice that the EDM constraint is usually the most
stringent bound on the amount of the CP violation in the theory. Therefore,
we consider $(\delta _{LL}^{d})_{23}$ as a promising candidate for large CP
violation which could be probed in rare B-decay processes, such as $%
B_{d}\rightarrow \phi K_{S}$.

$(\delta _{LL}^{d})_{23}$ is constrained by $b\rightarrow s\gamma $ because it
enters both
the gluino and chargino diagrams 
\footnote{In the SCKM base,
The contribution from chargino and up-type squark diagram involves
the factor $V^{\dagger}_{CKM} (\tilde{m}^2_u)_{LL} V_{CKM}$
\cite{Misiak:1997ei},
which is precisely the left-handed down type squark mass matrix
$(\tilde{m}^2_d)_{LL}$. Hence, $(\delta _{LL}^{d})_{23}$ also contributes to
$b\rightarrow s\gamma $ through chargino diagram.}.
Our numerical studies of $(\delta _{LL}^{d})_{23}$ are shown in Fig.~\ref%
{LLregion} and Fig.~\ref{LLregion-M2}. \footnote{We used
leading order running of Wilson coefficients for all the calculations
presented in this article.}
We see that it is possible to satisfy
the $b\rightarrow s\gamma $ bound and generate an interesting CP violating
effect in $B_{d}\rightarrow \phi K_{S}$ at the same time. More importantly,
while giving interesting low energy CP violation this scenario is \textit{not%
} constrained by the current EDM bound. 
For the entire parameter space shown in Fig. \ref{LLregion}, 
$|e\tilde{d}^C_s| < 8.5\times 10^{-26}ecm$.


\subsubsection{$(\protect\delta^d_{RR})_{23}|_{\Lambda_{GUT}}$}

\begin{figure}[h]
\center \epsfig{file=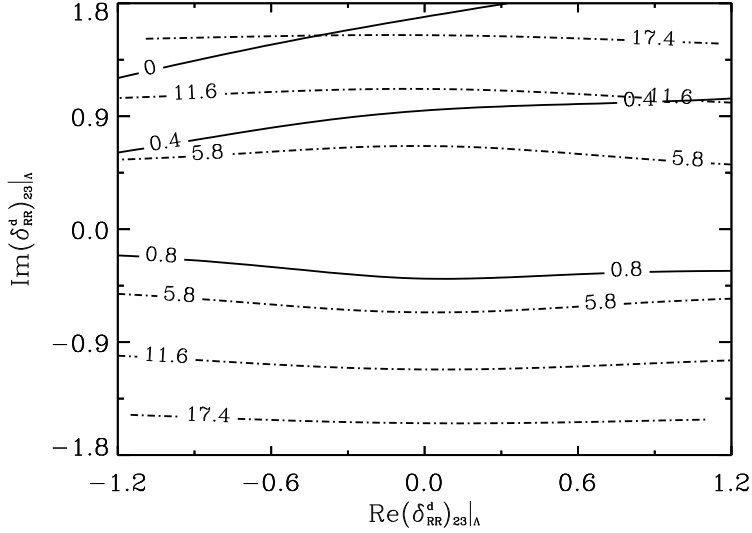,height=12cm, angle=90}
\caption{
This figure shows the correlation between $S_{\phi K_S}$ (solid lines)
and the $s$-quark CEDM (dashed lines in the unit of $10^{-25}ecm$). 
$(\delta^d_{RR})_{23}$ nearly does  not contribute to $b \to s \gamma$.
In view of the uncertainties associated with the $s$-quark CEDM,
we allow a  relaxation of a factor of 3 of the current experimental
bound.
\label{RRregion}
}
\end{figure}
$(\delta _{RR}^{d})_{23}$ could contribute to $S_{\phi K_{S}}$ through the $%
O_{8g}^{\prime }$ operator as $C_{8g}^{\prime }\propto (\delta
_{LR}^{d})_{33}^{\ast }(\delta _{RR}^{d})_{23}$. Since the right-handed
rotations are not constrained by the Standard Model CKM matrix, $(\delta
_{RR}^{d})_{23}$ is not in principle related to $(\delta _{RR}^{u})_{23}$.
Therefore, $(\delta _{RR}^{d})_{23}$ is not strongly constrained by $%
b\rightarrow s\gamma $ .

On the other hand, the mercury EDM strongly constrains the allowed CP
violation carried by $(\delta _{RR}^{d})_{23}$. The leading order
contribution is the combination $(\delta _{LR}^{d})_{23}(\delta
_{RR}^{d})_{23}^{\ast }$. The universal RGE contribution to $(\delta
_{LR}^{d})_{23}$ is suppressed. Combining this with the fact that RGE
evolution would not mix $(\delta _{RR}^{d})_{23}^{\ast }$ with $(\delta
_{LR}^{d})_{23}$, we conclude the that leading order contribution to the EDM
would not strongly constrain $(\delta _{RR}^{d})_{23}$. The next order
contribution is $(\delta _{LL}^{d})_{23}(\delta _{LR}^{d})_{33}(\delta
_{RR}^{d})_{23}^{\ast }\propto (\delta _{LL}^{d})_{23}C_{8g}^{\ast \prime }$.
As indicated in Table~\ref{rgedeltatable}, there is a universal RGE
contribution to $(\delta _{LL}^{d})_{23}\sim 0.01$. Hence, a large non-SM CP
violation in $S_{\phi K_{S}}$ from $C_{8g}^{\prime }$ would almost
generically imply a large contribution to the mercury EDM. Therefore, the
prospect of getting a large CP violation effect from $(\delta _{RR}^{d})_{23}$%
, such as proposed in Ref.\cite{Harnik:2002vs}, is necessarily constrained
by the mercury EDM bound, as indicated in Fig~\ref{RRregion}.

\subsubsection{$(\protect\delta^d_{LR})_{23}|_{\Lambda_{GUT}}$}

\begin{figure}[h!]
\begin{center}
\epsfig{file=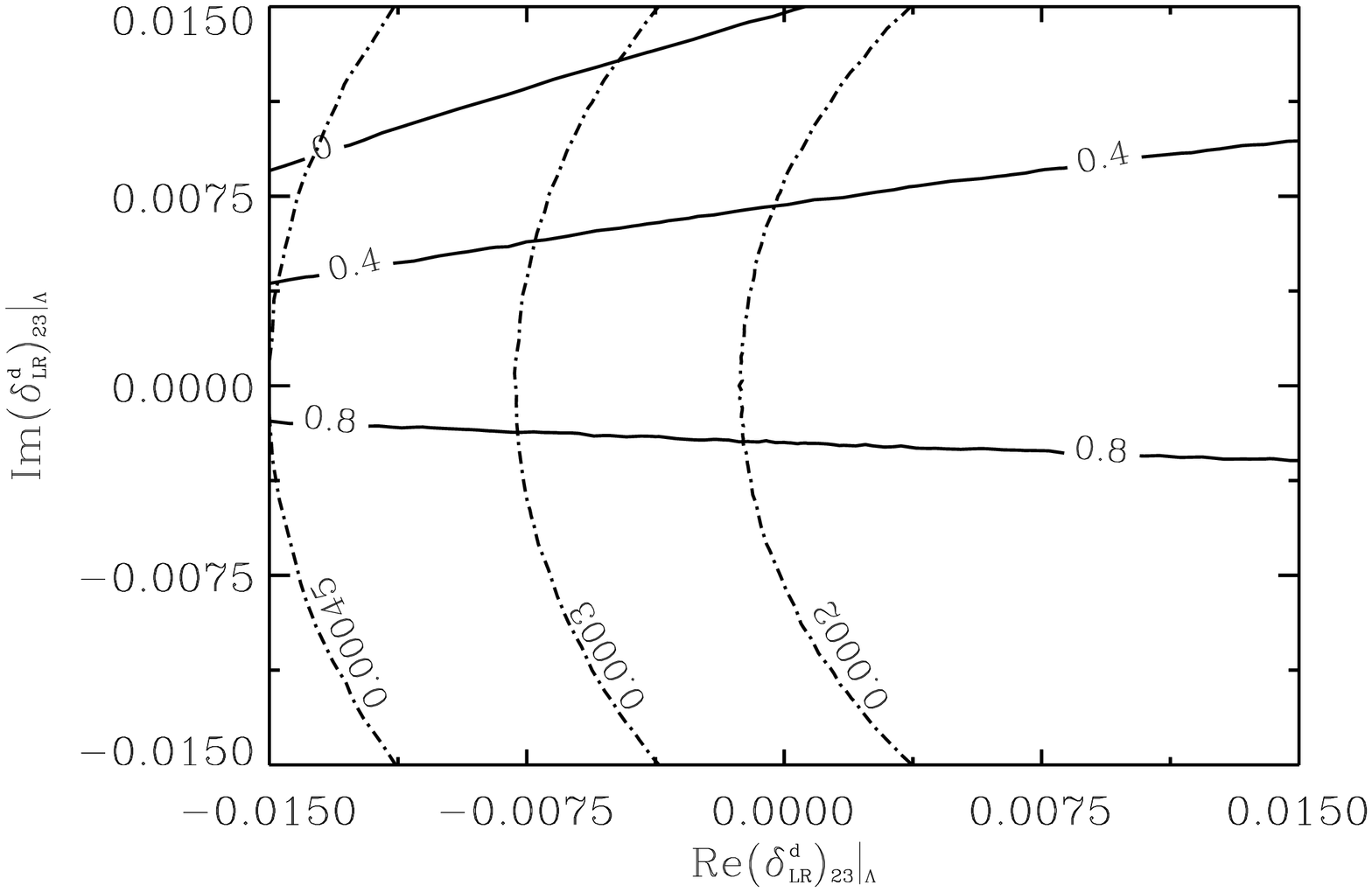,height=12cm, angle=90}
\caption{
The interplay between $S_{\phi K_S}$ (solid lines) and 
BR$(b\to s\gamma)$ (dashed lines) is shown.
We assume that $(\delta^u_{LR})_{23} = 0$ is not related to
$(\delta^d_{LR})_{23}$ at the input scale. As a consequence,
both BR$(b\to s\gamma)$ and the $s$-quark CEDM become less constraining
as opposed to the more usual case in Fig.(\ref{LRudregion})
where $A^{U,L}_i = A^{D,L}_i$ at the input scale with 
$\tilde{A}^f_{ij} = Y^f_{ij}A^{f,L}_i$ being the trilinear term ($f=U,D$).
\label{LRregion}
}
\end{center}
\end{figure}

\begin{figure}[h!]
\begin{center}
\epsfig{file=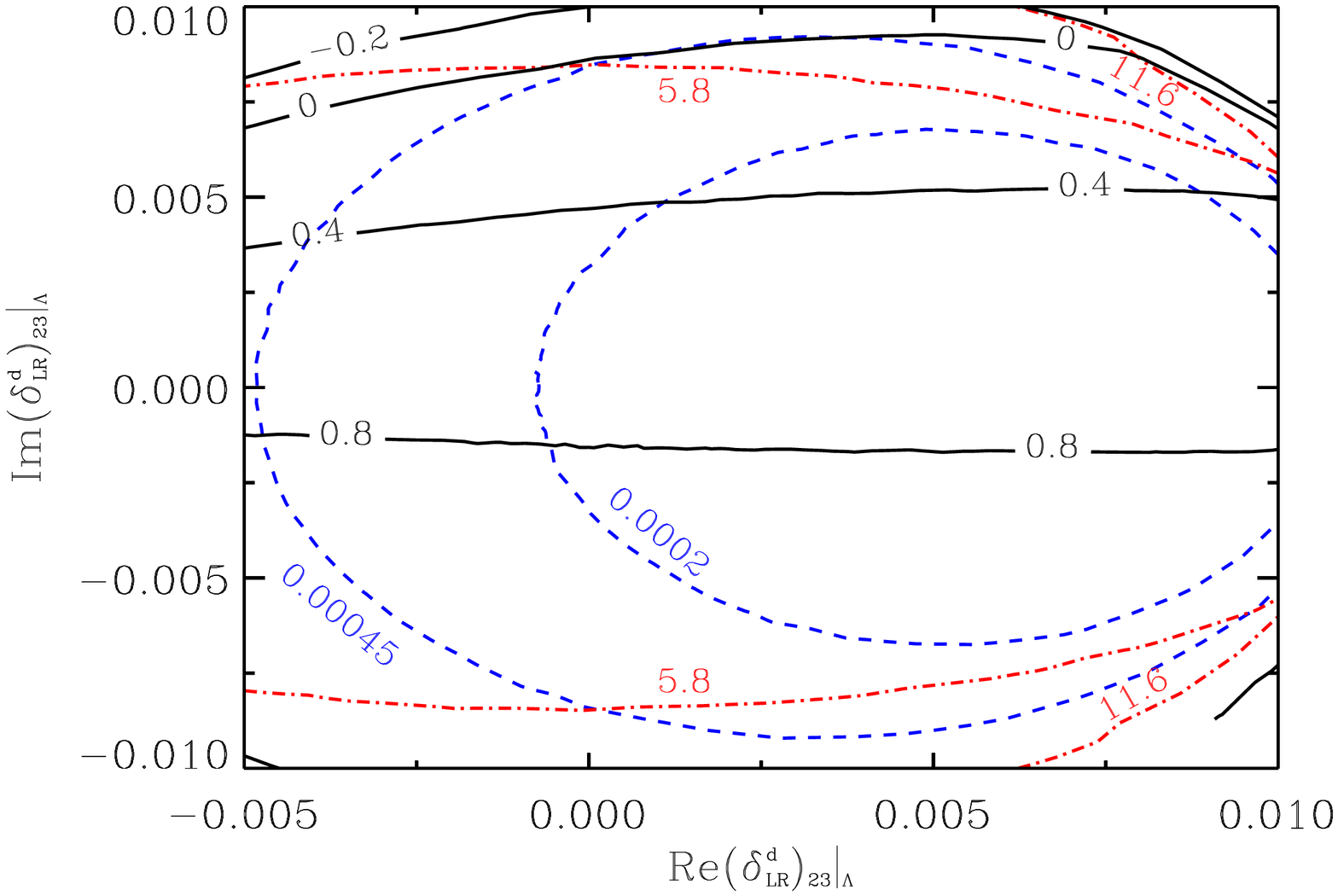,height=12cm, angle=90}
\caption{
This figure shows the interplay among $S_{\phi K_S}$ (black solid lines),
BR$(b\to s\gamma)$ (blue dash lines) and the $s$-quark chromo-EDM
(red dash-dotted lines, in the unit of $10^{-25}ecm$).
We assume
$A^{U,L}_i = A^{D,L}_i$ at the input scale, where
$\tilde{A}^f_{ij} = Y^f_{ij}A^{f,L}_i$ is the trilinear term ($f=U,D$).
\label{LRudregion}
}
\end{center}
\end{figure}

$(\delta _{LR}^{d})_{23}$ could contribute to $S_{\phi K_{S}}$ through the $%
O_{8g}$ operator as $C_{8g}\propto (\delta _{LR}^{d})_{23}$. It could give
rise to a large deviation from the Standard Model predictions.

The leading order contribution to the mercury EDM comes from combination $%
(\delta _{RR}^{d})_{23}^{\ast }(\delta _{LR}^{d})_{23}\propto (\delta
_{RR}^{d})_{23}^{\ast }C_{8g}$. It does not strongly constrain $(\delta
_{LR}^{d})_{23}$ since RGE running would not induce $(\delta
_{RR}^{d})_{23}^{\ast }$ either from $(\delta _{LR}^{d})_{23}$ or from
universal contributions. The next order contribution comes from the
combination $(\delta _{LR}^{d})_{23}(\delta _{LR}^{d})_{33}^{\ast }(\delta
_{LR}^{d})_{32}$. The universal contribution to $(\delta _{LR}^{d})_{32}$ is
highly suppressed. Since $(\delta _{LR}^{d})_{32}$ (approximately) does not
mix with $(\delta _{LR}^{d})_{23}$ in RGE running, the contribution to the
EDM is again suppressed at this order. Therefore, although $(\delta
_{LR}^{d})_{23}$ does feed strongly into other soft parameters such as $%
(\delta _{LL}^{d})_{23}$, CP violation in $(\delta _{LR}^{d})_{23}$ would
not be very strongly constrained by the mercury EDM.

$b\rightarrow s\gamma $ \ provides interesting constraints on $(\delta
_{LR}^{d})_{23}$. For a large class of models, $(\delta
_{LR}^{d})_{23}m_{t}/m_{b}\sim (\delta _{LR}^{u})_{23}$. $(\delta
_{LR}^{u})_{23}$ contributes to $b\rightarrow s\gamma $ \ through the
combinations $(\delta _{LR}^{u})_{23}(\delta _{LR}^{u})_{33}^{\ast }$ and $%
(\delta _{LR}^{u})_{23}^{\ast }(\delta _{LL}^{u})_{23}$. 

Numerical study of the $(\delta _{LR}^{d})_{23}$ scenario is shown in Fig.~%
\ref{LRudregion}. We also include a study of the scenario where $(\delta
_{LR}^{d})_{23}$ is not related to $(\delta _{LR}^{u})_{23}$, which is set to
zero at the high scale. The result is shown in Fig.\ref{LRregion}. 

\subsubsection{$(\protect\delta^d_{LR})_{32}|_{\Lambda_{GUT}}$}

\begin{figure}[h!]
\begin{center}
\epsfig{file=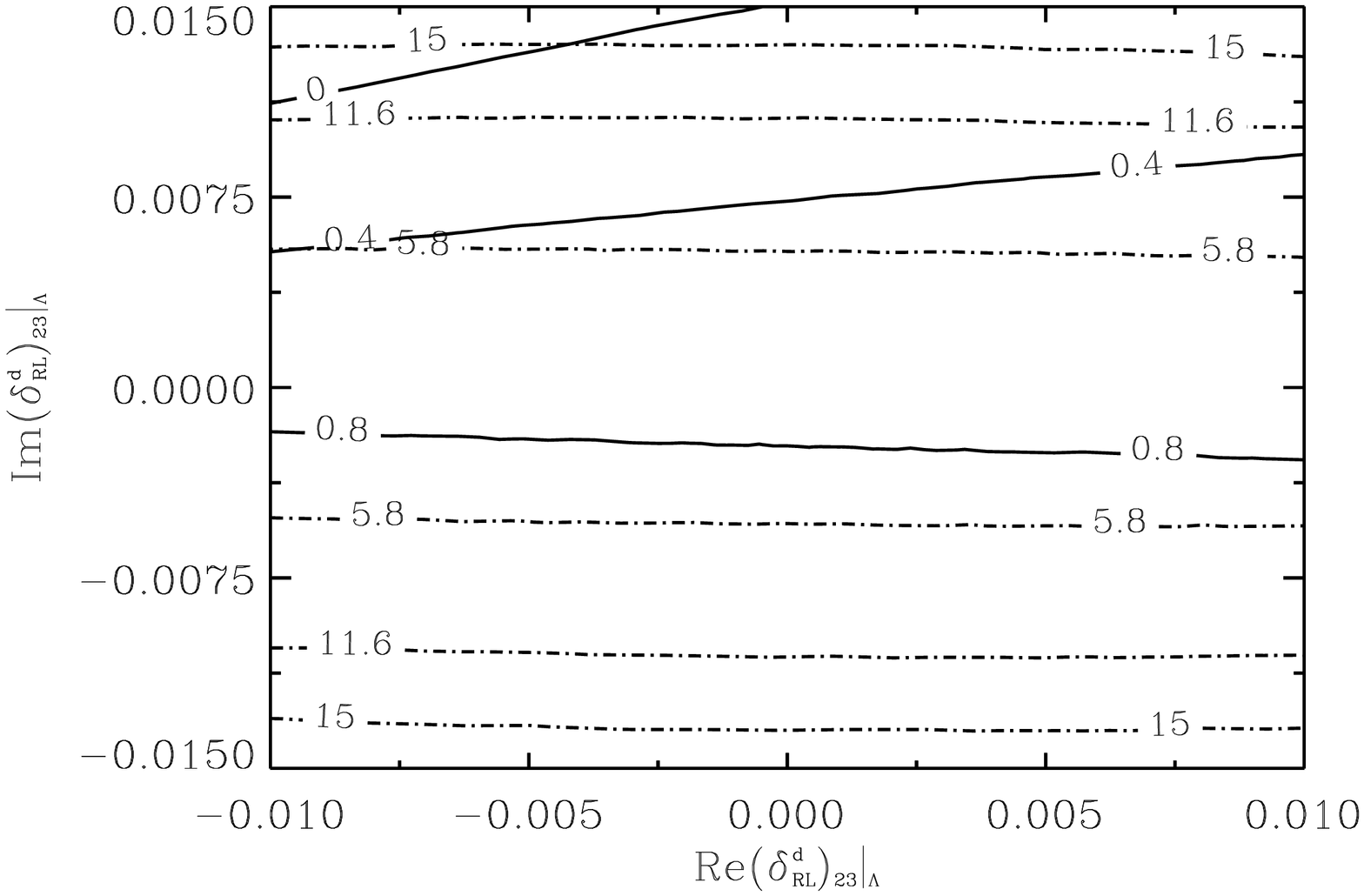,height=12cm, angle=90}
\caption{
The correlation between $S_{\phi K_S}$ (solid lines) and 
the $s$-quark CEDM (dashed lines, in the unit of $10^{-25}ecm$) is shown as 
a contour plot.
Br($B \to X_s\gamma$) imposes no constraint here, while the $s$-quark
CEDM is very constraining.
We relax the CEDM bound by a factor of $3$ to remind the reader of
its large uncertainties.
\label{RLregion}
}
\end{center}
\end{figure}

$(\delta _{LR}^{d})_{32}$ will contribute to $S_{\phi K_{S}}$ through the $%
O_{8g}^{\prime }$ operator as $C_{8g}^{\prime }\propto (\delta
_{LR}^{d})_{32}^{\ast }$. Since this is the first order in the mass insertion,
generically, $(\delta _{LR}^{d})_{32}$ could give rise to larger CP
violation in $C_{8g}^{\prime }$ than $(\delta _{RR}^{d})_{23}$. According to
Table~\ref{rgedeltatable}, an input scale $(\delta _{LR}^{d})_{32}$ also
feeds strongly into $(\delta _{RR}^{d})_{23}$ through RGE running.
Therefore, it could induce a larger contribution to $C_{8g}^{\prime }$
through $(\delta _{RR}^{d})_{23}$.

Since $(\delta _{RR}^{d})_{32}$ is not generically related to $(\delta
_{RR}^{u})_{32}$ at the high scale (see Section~\ref{modelsection}), $%
b\rightarrow s\gamma $ does not seriously constrain this scenario.

Since the CP violation entering $S_{\phi K_{S}}$ is from the operator $%
O_{8g}^{\prime }$, it is again strongly constrained by the $(\delta
_{LL}^{d})_{23}C_{8g}^{\ast \prime }$ contribution to the EDM. The
constraints are similar to those derived in the $(\delta _{RR}^{d})_{23}$
scenario. 
The numerical result is presented in Fig.~\ref{RLregion}. The conclusion is
that the prospect of CP violation in $(\delta _{LR}^{d})_{32}$ is highly
constrained by mercury EDM bound.

\section{Models of High Energy Flavor Structure}

\label{modelsection}

In this section, we study several different models of high energy flavor
structure. Our primary interest is in which models can have large low energy
CP violation beyond the Standard Model and which ones cannot, for generic
reasons. Given its considerable current interest, we will focus on the $%
B_{d}\rightarrow \phi K_{S}$ CP asymmetry. Our analysis could be generalized
to other low energy CP violating observables.

\subsection{General Estimates from Supergravity}\label{fl:sec:misu}

In this section, we give general estimates of various mass insertion
parameters using the general supergravity Lagrangian. To begin with, we
discuss a parameterization of the trilinears with rather
general assumptions about the supergravity induced soft terms.

In supergravity based models, the trilinear terms can be written as 
\begin{equation}
\widetilde{A}_{ij}^{U}=Y_{ij}^{U}A_{ij}^{U}\qquad \widetilde{A}%
_{ij}^{D}=Y_{ij}^{D}A_{ij}^{D}.  \label{fltri1}
\end{equation}%
Notice there is no matrix product in these two equations. It is in this
parameterization that the absolute values of $A_{ij}^{U,D}$ are at the order
of ${\mathcal{O}}(m_{3/2})$. 
One can certainly write trilinears as $\tilde{A}^{D}=Y^{D}\cdot
\bar{A}^{D}$ where a matrix product is used. In this form, it's not
guaranteed 
that every element of $\bar{A}$ is of order ${\mathcal{O}}(m_{3/2})$ and
the structure of $\bar{A}^{D}$ could be quite different from $A^{D}$
in eq.(\ref{fltri1}). Since we assume the MSSM Lagrangian arises from
supergravity theory, we will use the parameterization in eq.(\ref{fltri1}).

To get a good estimate of $\delta ^{d,LR}$, a reparameterization of eq.(\ref%
{fltri1}) will be useful. Take $A^{D}$ as an example. It can be written as 
\begin{equation}
A_{ij}^{D}=A_{0}^{D}+A_{i}^{D,L}+A_{j}^{D,R}+A_{11}^{D^{\prime }}\delta
_{i1}\delta _{j1}+A_{12}^{D^{\prime }}\delta _{i1}\delta
_{j2}+A_{21}^{D^{\prime }}\delta _{i2}\delta _{j1}+A_{22}^{D^{\prime
}}\delta _{i2}\delta _{j2}  \label{fltrip}
\end{equation}%
where 
\begin{eqnarray}
&&A_{0}^{D}=A_{33}^{D}  \label{fltridef} \\
&&A_{1}^{D,L}=A_{13}^{D}-A_{33}^{D}\qquad
A_{2}^{D,L}=A_{23}^{D}-A_{33}^{D}\qquad A_{3}^{D,L}=0  \nonumber \\
&&A_{1}^{D,R}=A_{31}^{D}-A_{33}^{D}\qquad
A_{2}^{D,R}=A_{32}^{D}-A_{33}^{D}\qquad A_{3}^{D,R}=0  \nonumber \\
&&A_{11}^{D^{\prime }}=A_{11}^{D}-A_{13}^{D}-A_{31}^{D}+A_{33}^{D}\qquad
A_{22}^{D^{\prime }}=A_{22}^{D}-A_{23}^{D}-A_{32}^{D}+A_{33}^{D}  \nonumber
\\
&&A_{12}^{D^{\prime }}=A_{12}^{D}-A_{13}^{D}-A_{32}^{D}+A_{33}^{D}\qquad
A_{21}^{D^{\prime }}=A_{21}^{D}-A_{23}^{D}-A_{31}^{D}+A_{33}^{D}  \nonumber
\end{eqnarray}%
If Yukawas are hierarchical, terms with a prime on the RHS of
eq.(\ref{fltrip}) have to multiply
small elements of the Yukawa matrix to give the corresponding elements in
the trilinears $\widetilde{A}^{D}$ . Thus under the assumption that all $%
A_{ij}$ are of order the gravitino mass and the Yukawas are hierarchical,
these terms are suppressed by at least $m_{s}/m_{b}\approx \lambda ^{2}$
where $\lambda =0.22$ is the Cabbibo angle. Hence we can safely neglect
them. Keeping this in mind, one has 
\begin{eqnarray}
\widetilde{A}_{ij}^{U} &\approx
&Y_{ij}^{U}(A_{0}^{U}+A_{i}^{U,L}+A_{j}^{U,R}) \\
\widetilde{A}_{ij}^{D} &\approx
&Y_{ij}^{D}(A_{0}^{D}+A_{i}^{D,L}+A_{j}^{D,R})  \nonumber  \label{fltrid}
\end{eqnarray}%
In most models, $A_{i}^{U,L}=A_{i}^{D,L}\equiv A_{i}^{Q}$ since $u_{L}^{i}$
and $d_{L}^{i}$ are in the same $SU(2)$ doublet. Under this assumption,
the above equations reduce to  
\begin{eqnarray}
\widetilde{A}_{ij}^{U} &=&Y_{ij}^{U}(A_{0}^{U}+A_{i}^{Q}+A_{j}^{U,R})
\label{fltrikv} \\
\widetilde{A}_{ij}^{D} &=&Y_{ij}^{D}(A_{0}^{D}+A_{i}^{Q}+A_{j}^{D,R}) 
\nonumber
\end{eqnarray}%
The authors of \cite{Masiero:2000rg} showed that trilinears of many SUSY
breaking models follow this parameterization even without the assumption of
hierarchical Yukawa couplings. Thus eq.(\ref{fltrikv}) can be applied to a
wide range of models.

To relate the trilinear terms to observables, we need to rotate $\widetilde{A%
}$ to the SCKM base. To be precise, let's define the down-type squark mass
insertion(MI) as 
\begin{equation}
\delta _{ij}^{d,LR}=\frac{(K_{L}^{D}\widetilde{A}^{D}K_{R}^{D\dag
})_{ij}v_{d}}{(\tilde{m}_{LL}^{D})_{ii}{(\tilde{m}_{RR}^{D})_{jj}}}
\label{flMIdef}
\end{equation}%
where $v_{d}$ is the vev of the down type higgs and matrices $K^{D}$ are the
rotation matrices that diagonalize Yukawa matrices. Among these $\delta _{ij}^{d,LR}$%
s, the most interesting ones are $\delta _{23}^{d,LR}$ and $\delta
_{32}^{d,LR}$ since they are directly related to the $B\rightarrow \phi
K_{S} $ signal. (Notice that $\delta _{ij}^{d,RL}=\delta _{ji}^{d,LR\ast }$ so
we only need to study $\delta _{ij}^{d,LR}$.) By using trilinears as written
in eq.(\ref{fltrid}), we obtain 
\begin{eqnarray}
\delta _{23}^{d,LR} &=&\frac{1}{(\tilde{m}_{LL}^{D})_{22}{(\tilde{m}%
_{RR}^{D})_{33}}}\times \\
&&\left( m_{b}(A_{3}^{D,L}-A_{2}^{D,L})(K_{L}^{D})_{23}(K_{L}^{D\ast
})_{33}+m_{b}(A_{1}^{D,L}-A_{2}^{D,L})(K_{L}^{D})_{21}(K_{L}^{D\ast
})_{31}\right.  \nonumber \\
&&\left. +m_{s}(A_{3}^{D,R}-A_{2}^{D,R})(K_{R}^{D})_{23}(K_{R}^{D\ast
})_{33}+m_{s}(A_{1}^{D,R}-A_{2}^{D,R})(K_{R}^{D})_{21}(K_{R}^{D\ast
})_{31}\right)  \nonumber
\end{eqnarray}%
Using the fact $m_{s}\ll m_{b}$,
$(K_{L}^{D})_{23}(K_{L}^{D\ast})_{33}\ll (K_{L}^{D})_{21}(K_{L}^{D\ast
})_{31}$
and the definitions in eq.(\ref{fltridef})
we simplify the above formula to 
\begin{equation}
\delta _{23}^{d,LR}\approx \frac{%
m_{b}(-A_{2}^{D,L})(K_{L}^{D})_{23}(K_{L}^{D\ast })_{33}}{m_{\tilde{q}}^{2}}=%
\frac{m_{b}(A_{33}^{D}-A_{23}^{D})(K_{L}^{D})_{23}(K_{L}^{D\ast })_{33}}{m_{%
\tilde{q}}^{2}}  \label{flLR23}
\end{equation}%
Using the same approximation, one also gets 
\begin{equation}
\delta _{32}^{d,LR}\approx \frac{%
m_{b}(-A_{2}^{D,R})(K_{R}^{D})_{33}(K_{R}^{D\ast })_{23}}{m_{\tilde{q}}^{2}}=%
\frac{m_{b}(A_{33}^{D}-A_{32}^{D})(K_{R}^{D})_{33}(K_{R}^{D\ast })_{23}}{m_{%
\tilde{q}}^{2}}  \label{flLR32}
\end{equation}%
Eq.(\ref{flLR23},\ref{flLR32}) give a quite good estimate of $\delta
_{23}^{d,LR}$ and $\delta _{32}^{d,LR}$ at the high scale. They suggest that
by increasing the splitting of trilinears between the 2nd and 3rd family
and/or increasing specific elements in the mixing($K^{D}$), these two MIs
could be enhanced.

Next, we make numerical estimates of individual mass insertion parameters.

\subsubsection{$\protect\delta^{d,LR}_{23}$}

The general form is given in eq.(\ref{flLR23}). The left-handed rotation $%
K_{L}^{D}$ is constrained not only by unitarity, but also by the CKM matrix: 
\begin{equation}
V_{23}^{CKM}=(K_{L}^{U})_{21}(K_{L}^{D\ast
})_{31}+(K_{L}^{U})_{22}(K_{L}^{D\ast })_{32}+(K_{L}^{U})_{23}(K_{L}^{D\ast
})_{33}=\lambda ^{2}
\end{equation}%
where $\lambda =0.22$ is the Cabibbo angle. If there are no large
cancellations among the 3 terms in the middle part of the above formula, one
expects that each of them is less than or around $\lambda ^{2}\approx 0.04$%
. This suggests $(K_{L}^{D})_{23}$ is much less than 1 and implies a large
suppression on $\delta _{23}^{d,LR}$. To avoid this suppression, one has to
assume both of the last two terms in the above equation are large and they
somehow cancel each other to get a small number $\lambda ^{2}$. This
could happen in models with democratic Yukawa couplings. If this
cancellation indeed happens, then from the unitarity constraint: 
\begin{equation}\label{fl:eq:ckmc}
|(K_{L}^{D})_{13}|^{2}+|(K_{L}^{D})_{23}|^{2}+|(K_{L}^{D})_{33}|^{2}=1
\end{equation}%
the maximum mixing we can have is $|(K_{L}^{D})_{23}(K_{L}^{D\ast
})_{33}|\approx 0.5$. In gravity-mediated SUSY breaking models, it's natural
to assume that both $A^{D}$ and $m_{\tilde{q}}$ are of order ${\mathcal{O}}%
(m_{3/2})$. Thus one gets $\delta _{23}^{d,LR}\approx 0.5\times m_{b}/m_{3/2}$%
. Suppose $m_{3/2}\approx 200$GeV, then $\delta _{23}^{d,LR}\approx {\mathcal{O%
}}(0.01)$.

This estimation is made at the high scale. To make contact with observables,
one must take the RGE running effect into account. The main RGE effect is to
enhance the diagonal terms in the squark mass matrix due to the gluino
contribution. The off-diagonal terms don't run much\cite{Ross:2002mr}.
Thus in eq.(\ref{flLR23}%
) and eq.(\ref{flLR32}), only the denominators are significantly affected by
the RGE running. The RGE running of the diagonal squark masses is
approximately 
\begin{eqnarray}
m_{\tilde{q}}^{2}|_{W} &\approx &6m_{1/2}^{2}+m_{0}^{2} \\
&=&6(\sqrt{3}m_{3/2})^{2}+m_{3/2}^{2}  \nonumber \\
&=&19m_{3/2}^{2}  \nonumber
\end{eqnarray}%
On the second line we assumed a dilaton dominated SUSY breaking scenario: $%
m_{1/2}=\sqrt{3}m_{3/2}$ and $m_{0}=m_{3/2}$. In the above formula,
$m_{1/2}$ and $m_0$ on the RHS
should take their high scale values. This formula shows
the low scale value of $\delta _{23}^{d,LR}$ will get a factor of 19
suppression from its high scale value in the dilaton dominated SUSY breaking
scenario, or a factor of 7 if one assumes $m_{1/2}\approx m_{0}\approx
m_{3/2}$. Thus the natural value of $\delta _{23}^{d,LR}$ is too small to
give a large deviation from the SM for the
$B_{d}\rightarrow \phi K_{S}$ process.

To compensate the RGE suppression, one needs to increase the splitting
between $A_{23}^{D}$ and $A_{33}^{D}$. For example, if we take $%
A_{23}^{D}=-A_{33}^{D}=4m_{3/2}$, there will be a factor of 8 enhancement in
the numerator of eq.(\ref{flLR23}). This allows $\delta _{23}^{d,LR}|_{W}$ to
be of order $0.01$ which could give rise to large CP violation in 
$B_{d}\rightarrow \phi K_{S}$ if we assume the cancellation happens 
in eq.(\ref{fl:eq:ckmc}).

From the discussion above, we see that in order to use a large $\delta^{d,LR}_{23}$
generating CP asymmetry in the $B_d \to\phi K_S$ process, the following
conditions should be satisfied:
\begin{itemize}
\item Large mixing in $K^D_L$: $|(K^D_L)_{23} (K^D_L)^*_{33}|\sim 0.5$
\item Large splitting between $A^D_{23}$ and $A^D_{33}$.
\end{itemize}
As we discussed, to satisfy the first condition, a large fine-tuning may be
required. 

\subsubsection{$\protect\delta^{d,LR}_{32}$.}

\label{fl:sec:LR32} The basic formula for $\delta _{32}^{d,LR}$ is shown in
eq.(\ref{flLR32}). The numerical estimate is similar to the $\delta
_{23}^{d,LR}$ case and there are two conditions to be satisfied if one uses $%
\delta _{32}^{d,LR}|_{\Lambda }$ to generate a large non-SM CP violation:
\begin{itemize}
\item Large splitting between $A^D_{32}$ and $A^D_{33}$: ($%
A^D_{32}-A^D_{33})\sim 8m_{3/2}$
\item Large right-handed mixing 
$|(K^D_R)_{23}^* (K^D_R)_{33}|\sim 0.5$ \label{fllr32con}.
\end{itemize}
Unlike $(K^D_L)_{23}$, there is no CKM constraints on $(K^D_R)_{23}$.



\subsubsection{Double Mass Insertions from $\protect\delta^{d,LL}_{23}$ and $%
\protect\delta^{d,RR}_{23}$.}

\label{fl:sec:DMI} 
At the weak scale, both single MI and double MI can give significant
contribution to $B\rightarrow \phi K_{S}$ process. In the MI approximation,
a double MI: $\delta _{33}^{d,LR}\times \delta _{32}^{d,RR}$ has an effect
similar to a single $\delta _{32}^{d,LR}$ MI, and 
$\delta _{33}^{d,LR}\times\delta _{23}^{d,LL}$ 
has an effect similar to $\delta _{23}^{d,LR}$.

The size of $\delta _{23}^{d,RR}$ can be estimated as follows. Assuming the
right-handed down-type squark mass matrix is diagonal in the gauge
eigenstates, we have 
\begin{eqnarray}
\delta _{32}^{d,RR} &=&\frac{(K_{R}^{D}\tilde{m}^{2}K_{R}^{D\dag })_{32}}{%
m_{3/2}^{2}} \\
&\approx &\frac{(\widetilde{m}_{D_{3}}^{2}-\widetilde{m}%
_{D_{2}}^{2})(K_{R}^{D})_{32}(K_{R}^{D\ast })_{33}}{m_{3/2}^{2}}  \nonumber
\\
&\sim &(K_{R}^{D})_{32}(K_{R}^{D\ast })_{33}\leq \frac{1}{2}  \nonumber
\end{eqnarray}%
Notice there is no fermion mass suppression and the only restriction on $%
K_{R}^{D}$ is unitarity. For $\delta _{23}^{d,LL}$, a similar estimate gives 
\begin{equation}
\delta _{23}^{d,LL}\approx (K_{L}^{D})_{23}(K_{L}^{D\ast })_{33}\sim \left\{ 
\begin{array}{cc}
\lambda ^{2} & \mbox{without cancellation} \\ 
0.5 & \mbox{with cancellation}%
\end{array}%
\right.
\end{equation}%
As in the single mass insertion case, the above estimates are at the high
scale. The RGE will induce large suppressions at the weak scale.


\subsection{Abelian Flavor Symmetry Models}

\label{fl:sec:5}

The discussion in the previous section is quite generic, not relying on any
flavor models. In this section, we study abelian flavor symmetry models and
discuss whether they can give rise to a non-SM CP violation.

Abelian flavor symmetry models are interesting since they give a nice
explanation for the observed hierarchical structure of fermion masses and
the CKM matrix. In the supergravity framework, the soft SUSY breaking terms
are related to the quantum numbers of the flavor symmetries\cite%
{Dudas:1996fe}. Therefore from a flavor symmetry model, we can calculate the
parameters of the soft supersymmetry breaking Lagrangian and learn whether
such models can give large deviations from low scale SM CP violation.

Suppose there is a $U(1)_{X}$ flavor gauge symmetry at the unification
scale. A field $\phi $ is a SM singlet but carries $U(1)_{X}$ charge: $%
X_{\phi }=-1$. Gauge invariance requires that the superpotential take the
form: 
\begin{eqnarray} \label{fl:sup}
W &=&\sum_{ij}Y_{ij}^{D}\theta (q_{i}+d_{j}+h_{d})\left( \frac{\phi }{M_{P}}%
\right) ^{q_{i}+d_{j}+h_{d}}Q_{i}D_{j}H_{d} \\
&+&\sum_{ij}Y_{ij}^{U}\theta (q_{i}+u_{j}+h_{u})\left( \frac{\phi }{M_{P}}%
\right) ^{q_{i}+u_{j}+h_{d}}Q_{i}U_{j}H_{u}  \nonumber 
\end{eqnarray}%
Here $M_{P}$ denotes the Planck scale. $q_{i}$, $d_{j}$ and $h_{d}$ are the $%
U(1)_{X}$ charges for $Q_{i}$, $D_{j}$ and $H_{d}$, respectively. $Y_{ij}^{D}
$ are some ${\mathcal{O}}(1)$ numbers. $\theta (x)=1$ for $x\geq 0$ and 0
otherwise. To get a hierarchical Yukawa matrix, $\left\langle \phi
\right\rangle /M_{P}$ should be a small number. We assume it is
approximately $\lambda \approx 0.22$. By choosing the $U(1)_{X}$ charges
correctly, one can generate hierarchical Yukawa matrices such as: 
\begin{equation}
Y^{U}\propto \left( 
\begin{array}{ccc}
\lambda ^{8} & \lambda ^{5} & \lambda ^{3} \\ 
\lambda ^{7} & \lambda ^{4} & \lambda ^{2} \\ 
\lambda ^{5} & \lambda ^{2} & 1%
\end{array}%
\right) \qquad Y^{D}\propto \left( 
\begin{array}{ccc}
\lambda ^{4} & \lambda ^{3} & \lambda ^{3} \\ 
\lambda ^{3} & \lambda ^{2} & \lambda ^{2} \\ 
\lambda ^{1} & 1 & 1%
\end{array}%
\right)   \label{flYd}
\end{equation}%
This set of Yukawas could give correct fermion masses and a correct CKM matrix.
Our discussion below doesn't depend on the detailed structure of the
Yukawas.

If the superpotential is coming from heterotic string theory, modular
invariance conditions should be satisfied. To specify these conditions, we
first write down the K\"ahler potential for the moduli fields: 
\begin{equation}
K=-\sum_{\alpha}\log(T_\alpha+T_\alpha^*)
\end{equation}
Here we denoted both T-type and U-type moduli fields collectively by $%
T_\alpha$. For each matter field $\Phi$, we denote the modular weights as $%
n^\alpha_{\Phi}$, corresponding to $T_\alpha$. In the superpotential, $%
Y^{U}_{ij}$($Y^{D}_{ij}$) may also depends on $T_\alpha$ and have modular
weights: $n_{U,ij}^\alpha$($n_{D,ij}^\alpha$). To keep the theory modular
transformation invariant, the following conditions should be satisfied\cite%
{Dudas:1996fe}: 
\begin{equation}
(q_i+d_j+h_d)n^\alpha_{\phi}+n^\alpha_{Q_i}+n^\alpha_{D_j}+
n^\alpha_{H_d}+n^\alpha_{D,ij}+1=0  \label{flmodcon}
\end{equation}

In the following, we would like to argue that in this model, it's difficult to
give a large beyond the SM contribution to the \bphik process. 
From the discussion of the previous section, we know that to get a large $%
\delta^{d,LR}_{32}$, $(K^D_{R})_{23}$ should be at ${\mathcal{O}}(1)$. This
requires 32 and 33 entries in $Y^D$ have similar magnitude, which implies
that they have the same powers of $\lambda$ due to eq.(\ref{fl:sup}), i.e. 
\begin{equation}
q_3+d_2+h_d=q_3+d_3+h_d\qquad \Rightarrow\qquad d_2=d_3.  \label{flcon1}
\end{equation}
In the flavor symmetry models, one usually assumes $Y_{ij}$ in eq.(\ref{flYd}%
) are ${\mathcal{O}}(1)$ and tries to explain the hierarchical structure by
using different powers of $\phi/M_P$. Under this assumption, $Y_{ij}$ should
be independent of moduli fields $T_\alpha$, or depend on $T_\alpha$ in
the same form. Therefore, all $n^\alpha_{D,ij}$ are equal.
Then using $d_2=d_3$, we have 
\begin{equation}
n^\alpha_{D_2}=n^\alpha_{D_3}  \label{flcon2}
\end{equation}

The two relations: $d_{2}=d_{3}$ and $n_{D_{2}}^{\alpha }=n_{D_{3}}^{\alpha }
$ have important implications for the soft terms. To calculate them we
assume a diagonal K\"{a}hler metric for the observable sector fields but
allow non-universality of the diagonal elements since different fields can
have different modular weights. The scalar masses and trilinear terms take
the form\cite{Dudas:1996fe}: 
\begin{eqnarray}
m_{ij}^{2} &=&m_{3/2}^{2}(1+\phi _{i}+3\cos ^{2}\theta \sum_{\alpha
}n_{i}^{\alpha }\Theta _{\alpha }^{2})\delta _{ij} \\
\widetilde{A}_{ij}^{D} &\equiv &A_{ij}^{D}Y_{ij}^{D}=m_{3/2}(-\sqrt{3}\sin
\theta +(q_{i}+d_{j}+h_{d}))Y_{ij}^{D}
\end{eqnarray}%
In the above equations, $\theta $ is the Goldstino angle and $\sum \Theta
_{\alpha }^{2}=1$. For the right-handed down type squarks, the second and
third generations have the same $U(1)_{F}$ charge and same modular weights,
so we have 
\begin{equation}
\widetilde{m}_{D_{2}}^{2}=\widetilde{m}_{D_{3}}^{2}.
\end{equation}%
Therefore, $\delta _{23}^{d,RR}=0$. For the trilinears, notice that they
satisfy the parameterization of eq.(\ref{fltrikv}). Thus eq.(\ref{flLR32})
should give a good estimate for $\delta _{32}^{d,LR}$ and we have 
\begin{eqnarray}
\delta _{32}^{d,LR} &\approx &\frac{m_{d}(A_{2}^{D}-A_{3}^{D})
(K_{R}^{D})_{33}(K_{R}^{D\ast })_{23}}
{m_{\tilde{q}%
}^{2}} \\
&=&\frac{m_{d}(d_{2}-d_{3})m_{3/2}(K_{R}^{D})_{33}(K_{R}^{D\ast })_{23}}
{m_{\tilde{q}}^{2}}  \nonumber \\
&=&0  \nonumber
\end{eqnarray}%
So in the large right-hand down type quark mixing case ($Y_{32}^{D}\approx
Y_{33}^{D}$), we have $\delta _{23}^{d,RR}\approx \delta _{32}^{d,LR}\approx
0$. By the same argument, one can show that in the large left-hand down
quark mixing case: $Y_{23}^{D}\approx Y_{33}^{D}$ , we have $\delta
_{23}^{d,RR}\approx \delta _{23}^{d,LR}\approx 0$. Therefore, in the large
mixing cases, it's difficult to generate large non-SM CP effects.

One can also try models with $\mathcal{O}(\lambda )$ suppressed mixing. For
example, $Y_{32}^{D}:Y_{33}^{D}=\mathcal{O}(\lambda )$. Then $%
(K_{R}^{D})_{23}\approx \lambda $ and according to the estimate in section %
\ref{fl:sec:LR32}, $|\delta _{32}^{LR}|$ will be much less than 0.01. In
this case, using the formula in section \ref{fl:sec:DMI}, we get $|\delta
_{23}^{RR}|\approx \lambda $ at the GUT scale. After running down to the
weak scale and therefore including a factor of about 7 RGE suppression, we
get $|\delta _{23}^{RR}|\approx \lambda /7\approx 0.03$. This MI contributes
to $B_{d}\rightarrow \phi K_{S}$ via a double mass insertion, which involves 
$\delta _{33}^{LR}$. To estimate $\delta _{33}^{LR}$, let's assume $m_{\tilde{q}%
}=\mu =600$ GeV and $\tan \beta =50$. Then we have 
\begin{equation}
\delta _{33}^{LR}\approx \frac{\overline{m_{b}}\mu \tan \beta }{m_{\tilde{q}%
}^{2}}\approx 0.24
\end{equation}%
where $\overline{m_{b}}$ is the running b-quark mass. Thus we can estimate
the double mass insertion 
\begin{equation}
\delta _{23}^{RR}\times \delta _{33}^{LR}\approx 0.03\times 0.24\approx 0.007
\end{equation}%
Remember for the double mass insertion case, there is an extra $1/2$
suppression from the Taylor expansion. In addition, for this case, the
2nd derivative of the loop 
function is smaller than the 1st derivative for the region around
$m_{\tilde g}^2/m_{\tilde q}^2=1$.
Thus there is no large deviation from the SM for the CP
asymmetry in  the $B_{d}\rightarrow \phi K_{S}$ process.
Similar arguments apply to the $Y_{23}^{D}:Y_{33}^{D}=\mathcal{O}(\lambda )$
cases. Therefore, without fine-tuning some parameters, we conclude that in
the $\mathcal{O}(\lambda )$ suppressed mixing cases, it's also difficult to
obtain a large non-SM $B_{d}\rightarrow \phi K_{S}$ result.

In summary, for Abelian flavor symmetry models, under the assumptions:
\begin{itemize}
\item The coefficients $Y^D_{ij}$ and $Y^U_{ij}$ in eq.(\ref{fl:sup}) are
all ${\mathcal{O}}(1)$ and T-moduli independent
\item K\"{a}hler metrics are diagonal(and allowed to be non-universal)
for the matter fields
\end{itemize}
it is unlikely to get a large non-SM contribution to the CP asymmetry of $%
B_{d}\rightarrow \phi K_{S}$ process.

\subsection{Family Dependent K\"{a}hlar Potential}
In the previous section we saw that if a $U(1)_{F}$ flavor symmetry model is
assumed, it's difficult to give a large deviation from the SM for the
\bphik process. Therefore, in
this section, we won't specify a particular flavor symmetry model. Instead, we
take the Yukawas as given parameters. Hence there are no direct relations
between the Yukawas and the soft terms. Then using a family dependent (but
still diagonal) K\"{a}hler potential, we will show that it's possible
to get a large SUSY contribution
to the CP asymmetry in $B_{d}\rightarrow \phi K_{S}$ .

We first give a model which has large mixing between the 2nd and 3rd
generation right-handed down type quarks. By splitting $A_{2}^{D}$ and $%
A_{3}^{D}$, we obtain a large $\delta _{32}^{d,LR}$ and therefore a large
contribution to $B_{d}\rightarrow \phi K_{S}$. Then we give a model which
has large mixing between the 2nd and 3rd generation left-handed quarks. By
splitting $A_{2}^{Q}$ and $A_{3}^{Q}$, we have a large $\delta _{23}^{d,LR}$
and it also gives a large contribution to $B_{d}\rightarrow \phi K_{S}$. In
the large $\delta _{32}^{d,LR}$ case, the EDM bound puts strong constraints on
how big the SUSY contribution to $B_{d}\rightarrow \phi K_{S}$ can be, and
in the large $\delta _{23}^{d,LR}$ case $b\rightarrow s\gamma $ constrains it.

It's also worth pointing out that in both models, the source of the CP
violation phases reside in the Yukawas and all the SUSY breaking $F$ terms($%
F_S$ and $F_T$) are real. If one takes the MSSM as an effective field
theory, only the invariant phases, such as the SM CP violation phase $%
\delta_{KM}$ or SUSY CP violation phase in eq.(\ref{fltheta32}), are
physical phases so that it doesn't matter whether we put the SUSY phases in
the Yukawas or $F$ terms. But from the underlying theory point of view,
Yukawas and $F$ terms have different origins and represent different physics.
Thus it's quite interesting that in our models, the only \textit{source} of
CP violation is in $Y_u$ and $Y_d$ and this may have important impact on
string model building.

\subsubsection{Large $\protect\delta^{d,LR}_{32}$ case.}
In this model, we take the Yukawas as given parameters. They have the
following form  at the high scale: 
\begin{eqnarray} \nonumber
Y_{u} &=&a_{u}\times \mbox{diag}\{m_{u},m_{c},m_{t}\} \\
Y_{d} &=&a_{d}\times V_{CKM}\cdot \mbox{diag}\{m_{d},m_{s},m_{b}\}\cdot U 
\end{eqnarray}%
In these equations, $a_{u}$ and $a_{d}$ are normalization factors depending
on $\tan \beta $. $V_{CKM}$ and the quark masses should take their high
scale values. $U$ is a matrix which generate a right handed down-type
quark mixing: 
\begin{equation}
U=\left( 
\begin{array}{ccc}
1 & 0 & 0 \\ 
0 & \cos\omega & e^{i\phi }\sin\omega \\ 
0 & -e^{-i\phi}\sin\omega  & \cos\omega %
\end{array}%
\right)   \label{flur}
\end{equation}%
Notice if we don't have soft susy breaking terms, the phase $\phi $ in $U$
is not observable and can be rotated away by field redefinition. But with
soft terms, certain combinations of $\phi $ and phases in soft terms are
physical observables and can't be rotated away. Also notice we assume the
above Yukawas are for already canonically normalized fields.

To calculate the soft terms, we assume a mixed dilaton/moduli susy breaking
scenario and assume we are in the weakly coupled heterotic orbifold vacuum.
The soft terms take the form as \cite{Brignole:1997dp} 
\begin{eqnarray}
m_{1/2} &=&\sqrt{3}m_{3/2}\sin \theta e^{-i\gamma _{S}} \\
m_{i}^{2} &=&m_{3/2}^{2}(1+3\cos ^{2}\theta \vec{n}_{i}.\vec{\Theta
^{2}}) \\ \label{flIBM3}
A_{ijk} &=&-\sqrt{3}m_{3/2}(\sin \theta e^{-i\gamma _{S}}+\cos \theta
\sum_{\alpha =1}^{3}e^{-i\gamma _{\alpha }}\Theta _{\alpha } \\ \nonumber
&\times &\left[ 1+n_{i}^{\alpha }+n_{j}^{\alpha }+n_{k}^{\alpha }+(T_{\alpha
}+T_{\alpha }^{\ast })\partial _{\alpha }\log (Y_{ijk})\right] )  
\end{eqnarray}%
In these formulas, $i,j,k$ denote different MSSM fields. $\theta $ is the
goldstino angle and $\sum \Theta _{\alpha }^{2}=1$ where $\alpha =1,2,3$
correspond to diagonal T-moduli fields associated with 3 compactified
complex planes. For simplicity we assume only these moduli fields are
relevant for soft term calculations. The $\gamma _{S}$ and $\gamma _{\alpha }
$ are the phases for $F_{S}$ and $F_{T_{\alpha }}$ and we set them to be
zero, so all the CP violation sources are in the Yukawas. $n_{i}^{\alpha }$
are the modular weights of a field $i$ with respect to $\alpha -$moduli. They
are negative fractional numbers. For fields $\phi $ in the untwisted sector,
the modular weights are 
$n_{\phi }=(-1,0,0)\mbox{ or }(0,-1,0)\mbox{ or }(0,0,-1)$
depending on which complex plane field $\phi $ is on. Modular weights for
twisted fields are a little bit more complicated and actually we don't
need to know them in this model.

Generally, the Yukawas $Y_{ijk}$ depend on moduli fields so that the last
term in eq.(\ref{flIBM3}) is nonzero. Since we take the Yukawas as input
numbers, there derivatives respect to $T_\alpha$ can
not be determined. Thus  we have to assume their
contributions to the trilinears are small and neglect them. Then the above
formula for the trilinears satisfies the parameterization in eq. (\ref{fltrid}).

Using these equations we have 
\begin{eqnarray} \nonumber
\Delta^A_{23}\equiv A^D_2-A^D_3&=&-\sqrt{3}m_{3/2}\cos\theta\sum_{%
\alpha=1}^6 e^{-i\gamma_\alpha}
\Theta_{\alpha}(n_{D_2}^{\alpha}-n_{D_3}^{\alpha}) \\ \nonumber
&=&-\frac{m_{1/2}e^{i\gamma_S}}{\tan\theta}\sum_{\alpha=1}^6
e^{-i\gamma_\alpha} \Theta_{\alpha}(n_{D_2}^{\alpha}-n_{D_3}^{\alpha}) 
\end{eqnarray}
From the discussions in section \ref{fl:sec:misu} we learned that to get large 
$\delta^{d,LR}_{32}$, $\Delta^A_{23}$ should be large. Thus we need a small
goldstino angle $\theta$ which means a moduli dominated susy
breaking scenario is preferred. 

In this model, we use the following parameters:
\begin{equation}
m_{3/2}=420\mbox{GeV} \qquad \sin\theta=\frac{1}{\sqrt{7}} \qquad 
\omega=\frac{\pi}{4} \qquad \tan\beta=40.
\end{equation}
The relevant modular weights and $\Theta _{\alpha }$ are shown in Table \ref%
{fl:tabLR32}. 
\begin{table}[h]
\caption{$\Theta _{i}$ and modular weights for some particles. }
\label{fl:tabLR32}
\begin{center}
\begin{tabular}{|p{2cm}|p{2cm}|p{2cm}|p{2cm}|}
\hline
& $T_1$ & $T_2$ & $T_3$ \\ \hline
$\Theta_i$ & 0 & -$\frac{1}{\sqrt{2}}$ & $\frac{1}{\sqrt{2}}$ \\ \hline
$n_{D_1}$ & 0 & -1 & 0 \\ \hline
$n_{D_2}$ & 0 & -1 & 0 \\ \hline
$n_{D_3}$ & 0 & 0 & -1 \\ \hline
$n_{H_u}$ & 0 & -1 & 0 \\ \hline
$n_{H_d}$ & 0 & -1 & 0 \\ \hline
\end{tabular}%
\end{center}
\end{table}
We assume the other matter fields have family independent modular weights.
Therefore, all other fields
have family independent trilinear and scalar masses. Notice $n_{D_{2}}$ and $%
n_{D_{3}}$ are different which means they have different K\"{a}hler
potentials 
\begin{equation}
K=\frac{|D_{2}|^{2}}{T_{2}+T_{2}^{\ast }}+\frac{|D_{3}|^{2}}{%
T_{3}+T_{3}^{\ast }}+...
\end{equation}%
This difference will only show up in the soft terms if $\cos \theta \neq 0$.

\begin{figure}[h]
\center \epsfig{file=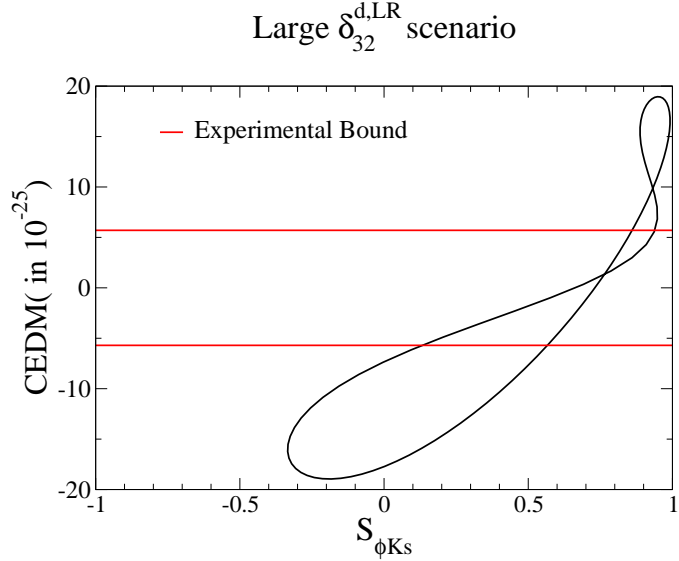,height=10cm, angle=-90}
\caption{CEDM vs. $S_{\protect\phi K_S}$ in the large $\protect\delta%
^{d,LR}_{32}$ scenario. We scan the phase $\phi$ in eq.(\ref{flur}) 
from 0 to $2\pi $. Two horizontal lines are the experimental bounds on
the s-quark CEDM. If we use the exact CEDM bound, the smallest $S_{\phi
K_S}$ we can get in this model is around 0.14. If we allow a factor 3
theoretical uncertainty on the CEDM prediction, the smallest  $S_{\phi
K_S}$ in this model is around -0.33. }
\label{fl:fig:LR32}
\end{figure}

Taking these parameters as high scale input, we use RGEs to run
the soft Lagrangian to the weak scale and calculate observables such as $%
S_{\phi K_{S}}$, the strange quark CEDM, $BR(b\rightarrow s\gamma )$, the
higgs mass, etc. We scan the phase $\phi $ in eq.(\ref{flur}) from 0 to $%
2\pi $. As explained before, for the large $\delta_{32}^{d,LR}$ case,
it's the EDM bound that is most difficult to satisfy. We 
show the correlation between predictions of $S_{\phi K_{S}}$ and $ed_{s}^{C}$
in figure \ref{fl:fig:LR32}. The upper bound on $|ed_{s}^{C}|$ is $5.8\times
10^{-25}e cm$. The theoretical prediction shown in this figure has a large
theoretical uncertainty. If we allow factor 3 theoretical uncertainty,
from this figure we find the smallest $S_{\phi K_{S}}$ is about -0.3. 
We checked that for this model, other experimental
constraints such as $BR(b\rightarrow s\gamma )$ and the higgs mass bound
are satisfied.

\subsubsection{Large $\protect\delta^{d,LR}_{23}$ case.}

\begin{figure}[h]
\center \epsfig{file=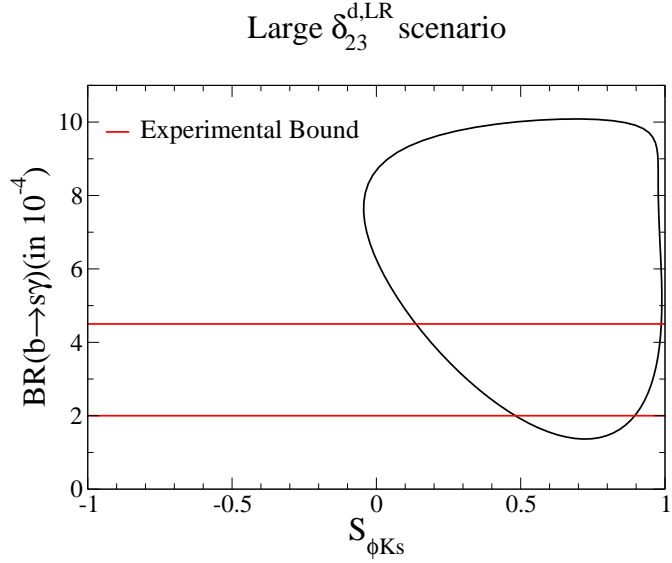,height=10cm, angle=-90}
\caption{ $BR(b\rightarrow s\protect\gamma )$ vs. $S_{\protect\phi K_{S}}$
in the large $\protect\delta _{23}^{d,LR}$ scenario. Two horizontal lines
are the bounds on $BR(b\rightarrow s\protect\gamma)$. The smallest  
$S_{\phi K_S}$ in this model without violating the 
$b\rightarrow s\protect\gamma$ constraint is around 0.13. 
 }
\label{fl:fig:LR23}
\end{figure}

To make this MI large, we need a large mixing of the 2nd and 3rd generation
left-handed quarks. Thus we use the following Yukawas: 
\begin{eqnarray}\nonumber
Y_{u} &=&a_{u}\times U\cdot\mbox{diag}\{m_{u},m_{c},m_{t}\} \\ \nonumber
Y_{d} &=&a_{d}\times U\cdot V_{CKM}\cdot\mbox{diag}\{m_{d},m_{s},m_{b}\} 
\end{eqnarray}%
The parameters are: 
\begin{equation}
m_{3/2}=360\mbox{GeV}\qquad \sin \theta =\frac{1}{\sqrt{7}}\qquad
\omega= \frac{\pi}{7}\qquad \tan\beta=24
\end{equation}%
The $\Theta _{\alpha }$ and modular weights are shown in Table \ref%
{fl:tabLR23}. 
\begin{table}[h]
\caption{$\Theta _{i}$ and modular weights for some particles }
\label{fl:tabLR23}
\begin{center}
\begin{tabular}{|p{2cm}|p{2cm}|p{2cm}|p{2cm}|}
\hline
& $T_1$ & $T_2$ & $T_3$ \\ \hline
$\Theta_i$ & 0 & -$\frac{1}{\sqrt{2}}$ & $\frac{1}{\sqrt{2}}$ \\ \hline
$n_{Q_1}$ & 0 & -1 & 0 \\ \hline
$n_{Q_2}$ & 0 & -1 & 0 \\ \hline
$n_{Q_3}$ & 0 & 0 & -1 \\ \hline
$n_{H_u}$ & 0 & -1 & 0 \\ \hline
$n_{H_d}$ & 0 & -1 & 0 \\ \hline
\end{tabular}%
\end{center}
\end{table}
Due to the different modular weights for $Q_{2}$ and $Q_{3}$, the trilinears 
$A_{2}^{Q}$ and $A_{3}^{Q}$ are split. (Notice $A^{Q}=A^{U,L}=A^{U,R}$.)
From eq.(\ref{flLR23}), we see that $\delta _{23}^{d,LR}$ will be large. We
assume modular weights for other particles are family independent. We
scan the phase $%
\phi $ in the $U$ matrix. As explained before, in the
large $\delta _{23}^{d,LR}$ case, the SUSY contribution to $S_{\phi K_{S}}$
is mainly constrained by $BR(b\rightarrow s\gamma )$. We show the correlations
between them in figure \ref{fl:fig:LR23}. From this figure, we see that
the smallest $S_{\phi K_{S}}$ 
we can get without violating the $BR(b\rightarrow s\gamma )$ bound is
about 0.13.  


\section{Conclusions}


Fundamental flavor physics is expected to manifest itself in the
flavor structure of the soft supersymmetry Lagrangian at the supersymmetry
breaking scale. In principle, there could be large CP violating parameters
in the soft Lagrangian. They could give rise to large deviations from
Standard Model prediction of the CP violating observables. One example is
the recent potential deviation in $S_{\phi K_S}$ of $B_d \to\phi K_S$ . On
the other hand, CP violation in the soft Lagrangian at high scale is
constrained by low energy observables such as EDM and $b\rightarrow s \gamma$.
In this paper, we give a model independent analysis of the constraints on
the high energy parameter space. Several interesting scenarios are
identified which both satisfy the experimental constraints and give rise to
large CP violation beyond the Standard Model. With the help of these results,
we further investigated classes of high energy flavor structure in the soft
parameters. We found that large classes of such models could not
produce large deviation from SM in processes such as $B_d \to\phi K_S$ and
satisfy the constraints at the same time. Finally, we presented several
scenarios where such a goal can be achieved. Thus such data could point
toward some classes of high scale theories as favored.
Further investigation along
those directions are clearly very interesting and important especially if
BELLE result on $S_{\phi K_S}$ of $B_d \to\phi K_S$ is confirmed.

\noindent{\bf Acknowledgement}

We would like to thank A. Kagan, P. Ko, B. Nelson, J.-h. Park, K. Tobe, 
L. Velasco-Sevilla, J. Wells and R.J. Zhang for helpful discussions. 
The work by HW is supported by the U.S. Department of Energy under
grant DE-FG02-91ER40681A29 (Task B). The research of the other authors
is supported in part by the U.S. Department of Energy.


\appendix

\section{RGE study of flavor parameters}

In this section, we analyze the RGE running of FCNC soft parameters. The
purpose of this study is to establish a systematic way to understand
semi-analytically the running of flavor off-diagonal parameters in the MSSM.
It provides an understanding complementary to that of the precise numerical
study (which is presented in Section~\ref{modelindepsection}) of the
connection between high and low scale flavor parameters. \ The techniques of
this appendix will be generally useful in future analysis.

In general, the running of a specific parameter in the SUSY Lagrangian is
basis-dependent since the RGEs depends on Yukawa couplings. The usual
practice is to write the RGE equations in the gauge eigenstate basis (so
that the gauge interactions are flavor diagonal). However, in this basis it
is not possible to give a systematic and model-independent study of the
running of the FCNC parameters due to the uncertainties of the Yukawa
couplings. Therefore, it is more useful to study the running in the so
called superCKM basis where there is a more direct connection of the low
energy observables and high energy parameters.

It is neither convenient, nor necessary, to diagonalize the Yukawa couplings
at each stage of the running. In order to get a systematic estimate of the
effect of the running, it is enough to run the RGE in the superCKM basis
defined by high energy input Yukawa couplings. More specifically, we
diagonalize the Yukawa matrices at the high energy input scale. This
diagonalization defines a high energy superCKM (HES) basis. We then run the
RGE equations in this basis. Now, in this basis, the RGE equations are
different from those written for the gauge eigenstates. However, it can be
shown that the only new parameters showing up the RGEs are the CKM matrix ($%
V_{CKM}$) obtained by diagonalizing the Yukawa matrices at the high scale.
Due to the facts (which we will justify) that the CKM matrix has only small
running \cite{Ramond:1993kv},
and the diagonal Yukawa couplings in this basis are proportional to
quark masses, it is thus possible to obtain reliable estimates of the
running.

We begin with Yukawa couplings. In the gauge eigenstate basis, the RGEs of
Yukawa couplings are\footnote{%
We are using the convention of Ref.~\cite{Chung:2003fi}.} 
\begin{eqnarray}
\frac{dY_{u}}{dt} &=&\frac{1}{16\pi ^{2}}[3Y_{u}Y_{u}^{\dagger
}Y_{u}+Y_{d}Y_{d}^{\dagger }Y_{u}+3Tr(Y_{u}^{\dagger }Y_{u})Y_{u}-(\frac{16}{%
3}g_{3}^{2}+6g_{2}^{2}+\frac{13}{15}g_{1}^{2})Y_{u}],  \nonumber \\
\frac{dY_{d}}{dt} &=&\frac{1}{16\pi ^{2}}[Y_{u}Y_{u}^{\dagger
}Y_{d}+3Y_{d}Y_{d}^{\dagger }Y_{u}+3Tr(Y_{d}^{\dagger }Y_{d})Y_{d}-(\frac{16%
}{3}g_{3}^{2}+3g_{2}^{2}+\frac{7}{15}g_{1}^{2})Y_{d}],  \label{rgeyukgauge}
\end{eqnarray}%
where we have suppressed all leptonic Yukawa couplings. We then rotate to
the HES basis by applying the transformation $Y\rightarrow V_{L}^{\ast
}YV_{R}^{T}$ on both sides of Eq.~\ref{rgeyukgauge}. We then obtain, in the
HES basis (we still denote the Yukawa coupling in this basis by $Y$) 
\begin{eqnarray}
\frac{dY_{u}}{dt} &=&\frac{1}{16\pi ^{2}}[3Y_{u}Y_{u}^{\dagger
}Y_{u}+V_{CKM}^{\ast }Y_{d}Y_{d}^{\dagger
}V_{CKM}^{T}Y_{u}+3Tr(Y_{u}^{\dagger }Y_{u})Y_{u}-(\frac{16}{3}%
g_{3}^{2}+6g_{2}^{2}+\frac{13}{15}g_{1}^{2})Y_{u}],  \nonumber \\
\frac{dY_{d}}{dt} &=&\frac{1}{16\pi ^{2}}[V_{CKM}^{T}Y_{u}Y_{u}^{\dagger
}V_{CKM}^{\ast }Y_{d}+3Y_{d}Y_{d}^{\dagger }Y_{u}+3Tr(Y_{d}^{\dagger
}Y_{d})Y_{d}-(\frac{16}{3}g_{3}^{2}+3g_{2}^{2}+\frac{7}{15}g_{1}^{2})Y_{d}],
\label{rgeyukhes}
\end{eqnarray}%
where the high energy CKM matrix $V_{CKM}$ is defined by $%
V_{u}^{L}V_{d}^{L\dagger }$.  In order to get an approximate solution to
these RGE equations, we make the following assumptions

\begin{enumerate}
\item The Yukawa matrices stay approximately diagonal in the running.
Therefore, approximately, individual diagonal entries run independently and
are proportional to quark masses.

\item The off-diagonal running terms, which are proportional to $V_{CKM}^{2}$%
, can be treated as a perturbation. The main effect of such a perturbation
is to generate off diagonal Yukawa couplings.

\item The CKM matrix does not run very much (or the running effect is
subleading). Therefore, we have a systematic expansion (in terms of $\lambda 
$) of the RGE effects.
\end{enumerate}

From these assumptions, we can solve for the additional mixing generated by
the running (the RGE running of the diagonal terms is well known and
dominated by SU(3) gauge coupling and third generation Yukawas) by
substituting diagonal Yukawa couplings in to Eq.~\ref{rgeyukhes} and
integrating approximately.

As a result, we have the following approximate solutions 
\begin{equation}
Y_u|_W \sim \left( 
\begin{array}{ccc}
y_u &  &  \\ 
& y_c &  \\ 
&  & y_t%
\end{array}%
\right) + \eta y_b^2 \left( 
\begin{array}{ccc}
& y_c \lambda^5 & y_t \lambda^3 \\ 
y_u \lambda^5 &  & y_t \lambda^2 \\ 
y_u \lambda^3 & y_c \lambda^2 & 
\end{array}%
\right),
\end{equation}

\begin{equation}
Y_{d}|_{W}\sim \left( 
\begin{array}{ccc}
y_{d} &  &  \\ 
& y_{s} &  \\ 
&  & y_{b}%
\end{array}%
\right) +\eta y_{t}^{2}\left( 
\begin{array}{ccc}
& y_{s}\lambda ^{5} & y_{b}\lambda ^{3} \\ 
y_{d}\lambda ^{5} &  & -y_{b}\lambda ^{2} \\ 
y_{d}\lambda ^{3} & y_{s}\lambda ^{2} & 
\end{array}%
\right) ,
\end{equation}%
where $\eta \sim |t_{EW}-t_{GUT}|/16\pi ^{2}\sim 0.2$. Since the diagonal
Yukawa couplings are proportional to the quark masses, the most significant
modification of the flavor mixing generated by RGE running are 
$\delta V_{13}^{L}\sim 0.1\lambda ^{3}$ and $\delta V_{23}^{L}\sim
0.1\lambda ^{2}$. We see that the $13$ and $23$ elements of the CKM matrix,
or a hierarchical $V^{L}$ proportional to the CKM matrix, could run about 10
percent. This is consistent with the assumption we made in solving the RGE
equations.

Next, we look at the running of the trilinears. In the HES basis, the RGE
equation for the trilinear $\tilde{A}_{d}$ is 
\begin{eqnarray}
\frac{d\tilde{A}_{d}}{dt} &=&\frac{1}{16\pi ^{2}}\{4\tilde{A}%
_{d}Y_{d}^{\dagger }Y_{d}+5Y_{d}^{\dagger }Y_{d}\tilde{A}%
_{d}+V_{CKM}^{T}Y_{u}Y_{u}^{\dagger }V_{CKM}^{\ast }\tilde{A}%
_{d}+2V_{CKM}^{T}\tilde{A}_{u}Y_{u}^{\dagger }V_{CKM}^{\ast }Y_{d}  \nonumber
\\
&+&\tilde{A}_{d}(3Tr[Y_{d}^{\dagger }Y_{d}]-\frac{16}{3}g_{3}^{2})+Y_{d}(%
\frac{16}{3}g_{3}^{2}M_{3}+6Tr[Y_{d}^{\dagger }\tilde{A}_{d}])\},
\end{eqnarray}%
where we suppressed subleading terms (such as terms proportional to
electroweak gauge couplings). First, we observe that the running of the
diagonal terms of the trilinears is almost always dominated by the term
proportion to the gluino mass (with the possible exception of
a large 3rd generation diagonal trilinear coupling). Intuitively, the
running of the off-diagonal terms are almost proportional to themselves.
Therefore, their running should not be very significant. To gain an
approximate understanding of the running, we could expand the RGE equations,
as we have done in the case of the Yukawa couplings, in terms of small
(off-diagonal) parameters such as $\lambda \sim 0.22$. We write the RGE
equation as 
\begin{equation}
16\pi^2\frac{d\tilde{A}_{d}}{dt}={\mathcal{A}}_{d}
\end{equation}%
where the righ-hand-side ${\mathcal{A}}_{d}$ is a $3\times 3$ matrix. In
terms of small parameters, the flavor off-diagonal entries of the last two
generations are 
\begin{eqnarray}
({\mathcal{A}}_{d})_{23} &\sim &-(y_{t}^{2}+y_{t}y_{b})\lambda ^{2}({\tilde{A%
}}_{d})_{33}+(y_{t}y_{b}+4y_{b}^{2})({\tilde{A}}_{d})_{23}+\lambda
y_{t}y_{b}({\tilde{A}}_{d})_{13}  \nonumber \\
({\mathcal{A}}_{d})_{32} &\sim &4\Delta y_{b}({\tilde{A}}%
_{d})_{33}+5y_{b}^{2}({\tilde{A}}_{d})_{32},
\end{eqnarray}%
where $\Delta \sim \eta y_{b}y_{t}^{2}\lambda ^{2}$, coming from term ${%
\tilde{A}}_{d}Y_{d}Y_{d}^{\dagger }$ in the RGE. The result of RGE running
of off diagonal entries of trilinear couplings is approximately ${\tilde{A}}%
_{d}=\eta {\mathcal{A}}_{d}$. Of course, it is always understood that the
entries in matrix ${\mathcal{M}}$ should be taking some appropriate
intermediate values. Useful estimates can be obtained from those
expressions. For example, we can derive that even without the presence of
off-diagonal terms in the trilinears, by RGE running, we will have\footnote{%
Notice that in our notation \cite{Chung:2003fi} , LR part of the squark mass
matrix corresponds to ${\tilde{A}}^{*}$} \footnote{%
In our estimation, we typically take the trilinear part to be the dominant
part in the LR sector of the squark mass matrix. This assumption could be
violated for the diagonal elements of LR sector, especially the down-sector
33 element, in the very large $\tan \beta $ and $\mu $ regime of the
parameter space. In that case, the estimates proportional ${\tilde{A}}%
_{33}^{d}$ (or mass insertions proportional to $(\delta _{LR}^{d})_{33}$
would be enhanced.)} 
\begin{equation}
(\delta _{LR}^{d})_{23}\sim \frac{v_{d}}{m_{{\tilde{q}}}}\frac{\eta \lambda
^{2}y_{t}^{2}({\tilde{A}}_{d}^{\ast })_{33}}{m_{{\tilde{q}}}}\sim 10^{-4},
\end{equation}%
and 
\begin{equation}
(\delta _{LR}^{d})_{32}\sim \frac{4\eta \Delta y_{b}v_{d}({\tilde{A}}%
_{d}^{\ast })_{33}}{m_{{\tilde{q}}}^{2}}\sim 10^{-5}
\end{equation}

Now we turn to consider the running of the soft masses. First, we consider
the running of the right-handed down-type soft mass parameters. The RGEs,
again in the HES basis, are 
\begin{equation}
{16\pi ^{2}}\frac{dm_{{\tilde{D}}}^{2}}{dt}={16\pi ^{2}}[2Y_{d}^{\dagger
}Y_{d}m_{\sD}^{2}+2m_{\sD}^{2}Y_{d}^{\dagger }Y_{d}+4Y_{d}^{\dagger }m_{\sQ%
}^{2}Y_{d}+4m_{H_{d}}^{2}Y_{d}^{\dagger }Y_{d}+4{\tilde{A}}_{d}^{\dagger }{%
\tilde{A}}_{d}-\frac{32}{3}g_{3}^{2}|M_{3}|^{2}].
\end{equation}
We could write the RHS as a $3\times 3$ matrix ${\mathcal{M}}_{\sD}$ and
expand it in terms of small parameters. For the last two generations, we
have approximately 
\begin{equation}
({\mathcal{M}}_{\sD})_{23}\sim 2\Delta y_{b}[(m_{\sD})_{33}^{2}+(m_{\sD%
})_{22}^{2}+2(m_{\sQ})_{33}^{2}]+4({\tilde{A}}_{d})_{32}^{\ast }({\tilde{A}}%
_{d})_{33}.
\label{rr23}
\end{equation}
Some important results can be derived from Eq.~\ref{rr23}. First, if we
begin at the input scale with a non-zero $({\tilde{A}}_{d})_{32}$, we would
induce a right-handed mixing term through RGE running 
\begin{equation}
(\delta _{RR}^{d})_{23}\sim \eta \frac{4({\tilde{A}}_{d})_{32}^{\ast }({%
\tilde{A}}_{d})_{33}}{m_{\sq}^{2}}
\sim 4\eta (\delta _{LR}^{d})_{32}^{\ast }
\frac{({\tilde{A}}_{d})_{33}}{v}\tan \beta 
\sim 100(\delta
_{LR}^{d})_{32}^{\ast }(\delta _{LR}^{d})_{33}.
\end{equation}%
Notice that a ${\tilde{A}}_{23}$ entry, on the other hand, does not generate
a large $RR$ mixing. This fact will be important in the search for viable
high scale models.

On the other hand, if there are only diagonal terms in the soft masses, the
RGE running could generate an off-diagonal mixing 
\begin{equation}
(\delta^d_{RR})_{23} \sim 2 \eta \Delta y_b [(m_\sD)_{33}^2+(m_\sD)_{22}^2+
2 (m_\sQ)_{33}^2]/m^2_{{\tilde{q}}} < 10^{-4}
\end{equation}
which is quite suppressed (comparing with the $LL $ case studied below ).

We also note that the mixing between $(\delta _{LL}^{d})_{23}$ and $(\delta
_{RR}^{d})_{23}$ is highly suppressed by second generation quark masses
and/or higher power of CKM mixing $\lambda $.

Finally, the RGE running of the soft masses of the left-handed squarks, in
the HES basis, are 
\begin{eqnarray}
\frac{dm_{{\tilde{Q}}_{d}}^{2}}{dt} &=&\frac{1}{16\pi ^{2}}%
\{V_{CKM}^{T}Y_{u}Y_{u}^{\dagger }V_{CKM}^{\ast }m_{{\tilde{Q}}_{d}}^{2}+m_{{%
\tilde{Q}}_{d}}^{2}V_{CKM}^{T}Y_{u}Y_{u}^{\dagger }V_{CKM}^{\ast
}+2V_{CKM}^{T}Y_{u}m_{{\tilde{U}}}^{2}Y_{u}^{\dagger }V_{CKM}^{\ast } 
\nonumber \\
&+&2m_{H_{u}}^{2}V_{CKM}^{T}Y_{u}Y_{u}^{\dagger }V_{CKM}^{\ast }+2V_{CKM}^{T}%
{\tilde{A}}_{u}{\tilde{A}}_{u}^{\dagger }V_{CKM}^{\ast }  \nonumber \\
&+&Y_{d}Y_{d}^{\dagger }m_{{\tilde{Q}}_{d}}^{2}+m_{{\tilde{Q}}%
_{d}}^{2}Y_{d}Y_{d}^{\dagger }+Y_{d}m_{{\tilde{D}}}^{2}y_{d}^{\dagger
}+2Y_{d}m_{{\tilde{D}}}^{2}Y_{d}^{\dagger }+2{\tilde{A}}_{d}{\tilde{A}}%
_{d}^{\dagger }-\frac{32}{3}g_{3}^{2}|M_{3}|^{2}\},
\end{eqnarray}%
which give 
\begin{eqnarray}
({\mathcal{M}}_{{\tilde{Q}}})_{23} &\sim &-y_{t}^{2}\lambda ^{2}[(m_{\sQ%
}^{2})_{33}+(m_{\sQ}^{2})_{22}+2(m_{\sU}^{2})_{33}+2m_{H_{u}}^{2}]  \nonumber
\\
&+&\Delta y_{b}[(m_{\sQ}^{2})_{33}+2(m_{\sD%
}^{2})_{33}+2m_{H_{d}}^{2}]+y_{b}^{2}(m_{\sQ}^{2})_{23}  \nonumber \\
&+&2({\tilde{A}}_{u})_{23}({\tilde{A}}_{u})_{33}^{\ast }+2({\tilde{A}}%
_{d})_{23}({\tilde{A}}_{d})_{33}^{\ast }+\lambda ({\tilde{A}}_{u})_{13}({%
\tilde{A}}_{u})_{33}^{\ast }.
\end{eqnarray}%
Therefore, in the absence of any off-diagonal terms, the RGE running will
generate a 
\begin{equation}
(\delta _{LL}^{d})_{23}\sim \eta y_{t}^{2}[(m_{\sQ}^{2})_{33}+(m_{\sQ%
}^{2})_{22}+2(m_{\sU}^{2})_{33}+2m_{H_{u}}^{2}]/m_{sq}^{2}\sim 4\eta
y_{t}^{2}=0.01.
\end{equation}%
Notice also, although a large ${\tilde{A}}_{23}$ does not generate a large $%
RR$ mixing, it will generate a sizable $LL$ mixing via RGE running.

Similarly, the mixing between $(\delta _{LL}^{d})_{23}$ and $(\delta
_{RR}^{d})_{23}$ is highly suppressed as well.

\end{document}